\documentclass{article}

\usepackage{graphicx}        % standard LaTeX graphics tool
\usepackage{multicol}        % used for the two-column index
\usepackage[bottom]{footmisc}% places footnotes at page bottom
\makeindex             % used for the subject index
\usepackage[matrix,arrow,ps]{xy}
\usepackage{graphicx}
\usepackage{subfigure}
\usepackage{amssymb,amsmath}
\usepackage{amsfonts}
\usepackage{graphicx}
\usepackage{type1cm}
\usepackage{eso-pic}
\usepackage{color}
\usepackage{makeidx}
\UsePSspecials{dvips}
\begin{document}

 \def\half{ {1\over 2} }
 \def\tab{ {\hskip 0.15 true in} }
 \def\vtab{ {\vskip 0.1 true in} }
 \def\htab{ {\hskip 0.1 true in} }
  \def\ntab{ {\hskip -0.1 true in} }
 \def\vtabb{ {\vskip 0.0 true in} }
 \def\blah{ {\vskip 0.1 true in} }
 \def\Order{\mbox{$\cal{O}$}}
 \def\vec#1{ {\overline {#1}} }
 \def\vecc#1{ {\overline {#1}} }
 \def\congruent{=}
\newcommand{\OO}{\mathbb{O}}
      \newcommand{\ba}{\left(\begin{array}}
      \newcommand{\ea}{\end{array}\right)}
      \newcommand{\beq}{ \begin{equation}}
      \newcommand{\eeq}{ \end{equation} }
       \newcommand{\IR}{\mathbb{R}}
        \newcommand{\IF}{\mathbb{F}}
        \newcommand{\IZ}{\mathbb{Z}}
           \newcommand{\II}{\mathbb{I}}
       \newcommand{\intrn}{\int_{\IR^n}}
       \newcommand{\intg}{\int_{G}}
        \newcommand{\sumn}{\sum_{i=1}^{n}}
        \newcommand{\sumN}{\sum_{i=1}^{N}}
 \newcommand{\IC}{\mathbb{C}}
       \newcommand{\IS}{\mathbb{S}}
       \newcommand{\sfrac}[2]{{\scriptstyle \frac{#1}{#2}}}
        \newcommand{\RR}{{\bf R}}
        \newcommand{\XX}{{\bf X}}
        \newcommand{\WW}{{\bf w}}
        \newcommand{\llangle}{\langle \hskip -0.02 true in \langle}
        \newcommand{\rrangle}{\rangle \hskip -0.02 true in \rangle}
\newcommand{\bfomega}{\mbox{\boldmath $\omega$ \unboldmath} \hskip -0.05 true in}
\newcommand{\gt}{X}
\newcommand{\JJ}{{\bf J}}
\newcommand{\xx}{{\bf x}}
\newcommand{\ff}{{\bf f}}
\newcommand{\yy}{{\bf y}}
\newcommand{\bb}{{\bf b}}
% \newcommand{\ba}{{\bf a}} % is this ever used ? If so, needs to be
% changed since \ba is defined as ``begin array'' above
\newcommand{\uu}{{\bf u}}
\newcommand{\dd}{{\bf d}}
\newcommand{\vv}{{\bf v}}
\newcommand{\ww}{{\bf w}}
\newcommand{\zz}{{\bf z}}
\newcommand{\qq}{{\bf q}}
\newcommand{\pp}{{\bf p}}
\newcommand{\grad}{{\rm grad}}
\newcommand{\rplus}{\mathbb{R}_{> 0}}
\newcommand{\rpluss}{\mathbb{R}_{\geq 0}}
\newcommand{\bfxi}{\mbox{\boldmath $\xi$ \unboldmath} \hskip -0.05 true in}
\newcommand{\bfsigma}{\mbox{\boldmath $\sigma$ \unboldmath} \hskip -0.05 true in}
\newcommand{\bftau}{\mbox{\boldmath $\tau$ \unboldmath} \hskip -0.05 true in}
\newcommand{\bfepsilon}{\mbox{\boldmath $\epsilon$ \unboldmath} \hskip -0.05 true in}
\newcommand{\bfphi}{\mbox{\boldmath $\phi$ \unboldmath} \hskip -0.05 true in}
\newcommand{\bfchi}{{\bf x}}
\newcommand{\bfdelta}{\mbox{\boldmath $\delta$ \unboldmath} \hskip -0.05 true in}
\newcommand{\bftheta}{\mbox{\boldmath $\theta$ \unboldmath} \hskip -0.05 true in}
\newcommand{\bfeta}{\mbox{\boldmath $\eta$ \unboldmath} \hskip -0.05 true in}
\newcommand{\bfmu}{\mbox{\boldmath $\mu$ \unboldmath} \hskip -0.05 true in}
\newcommand{\bfgamma}{\mbox{\boldmath $\gamma$ \unboldmath} \hskip -0.05 true in}
\newcommand{\bfalpha}{\mbox{\boldmath $\alpha$ \unboldmath} \hskip -0.05 true in}
\newcommand{\bflambda}{\mbox{\boldmath $\lambda$ \unboldmath} \hskip -0.05 true in}
\newcommand{\bfbeta}{\mbox{\boldmath $\beta$ \unboldmath} \hskip -0.05 true in}
\newcommand{\bfpi}{\mbox{\boldmath $\pi$ \unboldmath} \hskip -0.05 true in}
\newcommand{\bfnu}{\mbox{\boldmath $\nu$ \unboldmath} \hskip -0.05 true in}
\newcommand{\bfpsi}{\mbox{\boldmath $\psi$ \unboldmath} \hskip -0.05 true in}
\newcommand{\tx}{\tilde{X}}
\newcommand{\te}{\tilde{E}}
\newcommand{\xr}{\tilde{X}^r}
\newcommand{\xl}{\tilde{X}^l}

\newcommand{\emptysett}{\mbox{\O}}

\newcommand{\bfIto}{It\^{o} }
\newcommand{\Ito}{It\^{o} }
\newcommand{\Itoit}{{\it It\^{o}}}
\newcommand{\s}{\sin}
\newcommand{\al}{\alpha}
\newcommand{\be}{\beta}
\newcommand{\ga}{\gamma}
\newcommand{\ee}{{\bf e}}
\newcommand{\bea}{\begin{eqnarray*}}
\newcommand{\eea}{\end{eqnarray*}}
\newcommand{\beaq}{\begin{eqnarray}}
\newcommand{\eeaq}{\end{eqnarray}}
\newcommand{\bz}{{\bf 0}}
\newcommand{\askip}{\htab\htab {\rm and} \htab\htab}
\newcommand{\asskip}{\htab {\rm and} \htab}
\newcommand{\wskip}{\htab {\rm where} \htab}
\newcommand{\wwskip}{\htab\htab {\rm where} \htab\htab}
\newcommand{\cX}{\tilde{X}}

\newcommand{\dimm}{m}
\newcommand{\dinn}{n}
\newcommand{\plus}{+}

\newcommand{\intf}{\int_{-\infty}^{\infty}}

%% in discussion of Maurer-Cartan, am using omega, but might change
%% to xi later, so better to be flexible by defining these
\newcommand{\om}{\omega}
\newcommand{\Om}{\Omega}
\newcommand{\bom}{\bfomega}
\newcommand{\trip}{\|\hskip -0.015 true in |}
\newcommand{\bltrip}{\left\|\hskip -0.02 true in \left|}
\newcommand{\brtrip}{\right\|\hskip -0.02 true in \right|}
\newcommand{\com}{********}
\newcommand{\ggloss}{\index}
\newcommand{\authgloss}{\index}

\makeatletter
%\AddToShipoutPicture{%
%          \setlength{\@tempdimb}{.5\paperwidth}%
%            \setlength{\@tempdimc}{.5\paperheight}%
%            \setlength{\unitlength}{1pt}%
%            \put(\strip@pt\@tempdimb,\strip@pt\@tempdimc){%
%        \makebox(0,0){\rotatebox{45}{\textcolor[gray]{0.85}%
%        {\fontsize{3cm}{3cm}\selectfont{\copyright 2/11/09}}}}%
%            }%
%
%            \setlength{\@tempdimb}{.5\paperwidth}%
%            \setlength{\@tempdimc}{.7\paperheight}%
%            \setlength{\unitlength}{1pt}%
%            \put(\strip@pt\@tempdimb,\strip@pt\@tempdimc){%
%        \makebox(0,0){\rotatebox{45}{\textcolor[gray]{0.85}%
%        {\fontsize{3cm}{3cm}\selectfont{gregc@jhu.edu}}}}%
%            }%
%}

\makeatother

\title{Information-Theoretic Inequalities \\ on Unimodular Lie Groups}
\author{Gregory S. Chirikjian \\Department of Mechanical Engineering \\ Johns Hopkins University \\ gregc@jhu.edu}
\maketitle

\pagenumbering{arabic}
\setcounter{page}{1}

\begin{abstract}

Classical inequalities used in information theory such as those of de Bruijn, Fisher, and Kullback 
carry over from the setting of probability theory on Euclidean space to that of unimodular Lie groups. These are groups that posses integration measures that are invariant under left and right shifts, which means that even
in noncommutative cases they share many of the useful features of Euclidean space. In practical
engineering terms the rotation group and Euclidean motion group are the unimodular Lie groups of most interest,
and the development of information theory applicable to these Lie groups opens up the potential to study
problems relating to image reconstruction from irregular or random projection directions, information gathering in mobile robotics, satellite attitude control, and bacterial chemotaxis and information processing. Several definitions are extended from
the Euclidean case to that of Lie groups including the Fisher information matrix, and inequalities analogous to those in classical information theory are derived and stated in the form of fifteen small theorems.
In all such inequalities, addition of random variables is replaced with the group product, and the appropriate generalization of convolution of probability densities is employed.

\end{abstract}

\section{Introduction}

Shannon's brand of information theory is now more than six decades old, and some of the statistical methods developed by Fisher, Kullback, etc., are even older. Similarly, the study of Lie groups is now more than a
century old. Despite their relatively long and roughly parallel history,
surprisingly few connections appear to have been made between these two vast fields.
One such connection is in the area of ergodic theory \cite{adler, Arnolerg, Birkhofferg},
where the Boltzmann-Shannon entropy is replaced with
topological entropy \cite{halmos, khinchin, sinaient, Wienererg}. 
Ergodic theory developed in parallel with information theory
and remains an active area of research among mathematicians to the current day (see e.g., \cite{Zimmererg}). Both use concepts of entropy
(though these concepts are quite different from each other), and some common treatments have been given over the years (see e.g., \cite{billingsley}). However, it should be noted that some of the cornerstones of information theory such as the  de Bruijn inequality, Fisher information, Kullback-Leibler divergence, etc.,
do not carry over to ergodic theory.
And while connections between ergodic theory and Lie groups are quite strong, connections between
information theory and Lie groups are virtually nonexistent.
The goal of this paper is therefore to present a unified framework of ``information theory on Lie groups.''
As such, fifteen small theorems are presented that involve the structure and/or group operation of Lie groups.
Unlike extensions of information theory to manifolds, the added structure inherent in Lie groups allow us
to draw much stronger parallels with inequalities of classical information theory, such as those presented
in \cite{Cover, dembo}.

In recent years a number of connections have begun to emerge linking information theory, group theory, and geometry. A cross section of that work is reviewed here, and it is explained how the results of this paper are distinctly different from prior works.

In the probability and statistics literature, the statistical properties of random walks and limiting
distributions on
Lie groups has been studied extensively by examining the properties of iterated convolutions
\cite{Diaconis, grenander, Heyer, Shlosman1, Shlosman2, Shlosman3, Stromberg}. 
The goal in many of these works is to determine the form of the
limiting distribution, and the speed of convergence to it. This is a problem closely related to those
in information theory. However, to the author's knowledge concepts such as entropy, Fisher information,
Kullback-Leibler divergence, etc., are not used significantly in those analysis. Rather, techniques of
harmonic analysis (Fourier analysis) on Lie groups are used, such as the methods described in \cite{book, Gelfand, Gurarie, Miller, sugiura, Taylor, varadarajan, Vilenkin, zhelobenko}.
Indeed, to the best of the author's knowledge the only work that uses the concept and properties of
information-theoretic (as opposed to topological) entropy on Lie groups is that of Johnson and Suhov 
\cite{johnson1, johnson2}. Their goal was to use the Kullback-Leibler divergence between probability density functions
on compact Lie groups to study the convergence to uniformity under iterated convolutions, in analogy with
what was done by Linnik \cite{Linnik} and Barron \cite{Barron} in the commutative case. The goal of the present
paper is complementary: using some of the same tools, many of the major defined quantities and inequalities of (differential) information theory are extended from $\IR^n$ to the context of unimodular Lie groups, which form a broader class of Lie groups than compact ones.

The goal here is to define and formalize probabilistic and information-theoretic quantities that are currently
arising in scenarios such as robotics \cite{lee,Makadia, hughdw, Roussopoulos, thrun,wooramrobotica,Kutzer,wang08}, bacterial motion \cite{berg, infotaxis}, and parts assembly in automated manufacturing systems \cite{Boothroyd, CASE09, Kim,sanderson,SuLee}.
% , and the reconstruction of the biomolecular shape from experimental data \cite{wooramnewest}.
The topics of detection, tracking, estimation and control on Lie groups has been studied extensively over the past four decades. For example, see \cite{15brocket73, 15brock73, censi09, Duncan, 15jurd, Steele, book, wooramrobotica, wang08, mahony09a, mahony09b,Willsky} (and references therein). Many of these problems involve probability densities on the group of rigid-body motions. However, rather than focusing only on rigid-body motions, a general information theory on the much
broader class of unimodular Lie groups is presented here with little additional effort.

Several other research areas that would initially appear to be related to the present work have received
intensive interest.
For decades, Amari has developed the concept of information geometry \cite{Amari} in which the Fisher information
matrix is used to define a Riemannian metric tensor on spaces of probability distributions, thereby allowing those spaces to be viewed as Riemannian manifolds. This provides a connection between information theory and differential geometry. However, in information geometry, the probability distributions themselves (such as Gaussian distributions) are defined on a Euclidean space, rather than on a Lie group.

A different kind of connection between information theory and geometry has been established in the context
of medical imaging and computer vision in which probability densities on manifolds are analyzed using information-theoretic techniques \cite{Pennec11}. However, a manifold generally does not
have an associated group operation, and so there is no natural way to ``add'' random variables.

Relatively recently,
Yeung and coworkers have used the structure of finite groups to derive new inequalities for discrete information.
While this heavily involves the use of the theory of finite groups, the goal is to derive new inequalities for
classical information theory, i.e., that which is concerned with discrete information related to finite sets.
For example, see the work of Chan and Yeung \cite{ChanYeung, Chan1} and Zhang and Yueng \cite{ZhangYueng}.
Li and Chong \cite{LiChong} and Chan \cite{Chan1} have addressed the relationship between group homomorphisms and
information inequalities using the Ingleton inequality. In these works, the groups are discrete,
and the new inequalities that are derived pertain to classical informational quantities. In contrast, the
goal of the current presentation is to extend concepts from information theory to the case where variables
``live in'' a Lie group.

While on the one hand work that connects geometry and information theory exists, and on the other hand work that
connects finite-group theory and information theory exists, very little has been done along the lines of
developing information theory on Lie groups, which in addition to possessing the structure of differential
manifolds, also are endowed with group operations.
Indeed, it would appear that applications such as deconvolution on Lie groups \cite{myinvprob}
(which can be formulated in an information-theoretic context \cite{yazici, kim2}), and
the field of Simultaneous Localization and Mapping (or SLAM) \cite{thrun} have preceded
the development of formal information inequalities that take advantage of the Lie-group structure
of rigid-body motions.

This paper attempts to address this deficit with a two-pronged approach: (1) by collecting some known results
from the functional analysis literature and reinterpreting them in information-theoretic terms (e.g. Gross' log-Sobolev inequality on Lie groups);
(2) by defining information-theoretic quantities such as entropy, covariance and Fisher information matrix, and
deriving inequalities involving these quantities that parallels those in classical information theory.

The remainder of this paper is structured as follows: Section \ref{revliegr} provides a brief review
of the theory of unimodular Lie groups and gives several concrete examples (the rotation group, Euclidean
motion group, Heisenberg group, and special linear group). 
%The concept of probability density functions on unimodular Lie groups and associated definitions of mean and covariance are reviewed.
An important distinction between information
theory on manifolds and that on Lie groups is that the existence of the group operation in the latter case
plays an important role.  Section
\ref{infogroupsec} defines entropy and relative entropy for unimodular Lie groups and proves some of
their properties under convolution and marginalization over subgroups and coset spaces. The concept of the Fisher information matrix for probability densities on unimodular Lie groups is defined in
Section \ref{fisherinfoliegr} and several elementary properties are proven. This generalized concept of Fisher
information is used in Section \ref{deBruijngr} to establish the de Bruijn inequality for unimodular Lie groups.
Finally, these definitions and properties are combined with recent results by others on log-Sobolev inequalities
in Section \ref{logsoblie}.

\section{A Brief Review of Unimodular Lie Groups} \label{revliegr}

Rather than starting with formal definitions, examples of unimodular Lie groups are first introduced,
their common features are enumerated, and then their formal properties are enumerated.

\subsection{An Introduction to Lie Groups via Examples}

Perhaps one reason why there has been little cross-fertilization between the theory of Lie groups
and information theory is that the presentation styles in these two fields are very different.
Whereas Lie groups belong to pure mathematics, information theory emerged from engineering. Therefore,
this section reviews some on the basic properties of Lie groups from a concrete engineering perspective.
All of the groups considered are therefore matrix Lie groups.

\subsubsection{Example 1: The Rotation Group}

Consider the set of $3\times 3$ rotation matrices
$$ SO(3) = \{R \in \IR^{3\times 3} \,|\, RR^T = \II, {\rm det} R = +1\}. $$
Here $SO(3)$ denotes the set of special orthogonal $3\times 3$ matrices with real entries.
It is easy to verify that this set is closed under matrix multiplication and inversion.
That is, $R, R_1,R_2 \in SO(3) \,\Longrightarrow\, R_1 R_2, R^{-1} \in SO(3)$. Furthermore,
the $3\times 3$ identity matrix is in this set, and the associative law $R_1(R_2 R_3) = (R_1 R_2) R_3$
holds, as is true for matrix multiplication in general. This means that $SO(3)$ is a group, and is called
the special orthogonal (or rotation) group. Furthermore,
it can be reasoned that the nine independent entries in a $3\times 3$ real matrix are constrained by
the orthogonality condition $RR^T = \II$ to the point where a three-degree-of-freedom subspace remains.
(The condition ${\rm det} R = +1$ does not further constrain the dimension of this subspace, though it does limit
the discussion to one component of the space defined by the orthogonality condition).

It is common to describe the three free degrees of freedom of the rotation group using parametrizations such
as the ZXZ Euler angles:
\beq
R(\alpha,\beta,\gamma) = R_3(\alpha) R_1(\beta) R_3(\gamma)
\label{eulerangle}
\eeq
where $R_i(\theta)$ is a counterclockwise rotation about the $i^{th}$ coordinate axis.
Another popular description of 3D rotations
or the axis-angle parametrization
\beq
R(\vartheta,{\bf n}) = \II + \sin \vartheta N + (1-\cos\vartheta) N^2
\label{axisangle}
\eeq
where $N$ is the unique skew-symmetric matrix such that $N \xx = {\bf n} \times \xx$ for any $\xx \in \IR^3$,
and ${\bf n}$ is the unit vector pointing along the axis of rotation and $\times$ is the vector cross product.
The ``vee and hat'' notation
\beq
N^{\vee} = {\bf n} \,\,\,\,\,\, \Longleftrightarrow \,\,\,\,\,\, N = \hat{\bf n}
\label{vdefeq}
\eeq
is used to describe this relationship.
Here $\|{\bf n}\| = ({\bf n}\cdot{\bf n})^{\half} = 1$. It can be parameterized in spherical coordinates as
${\bf n} = {\bf n}(\phi,\theta)$, and so a parametrization of the form $R = R(\vartheta,\phi,\theta)$ results.
The angles $\vartheta,\phi,\theta$ are not the same as the Euler angles $\alpha,\beta,\gamma$.

The group $SO(3)$ is a compact Lie group, and therefore has finite volume. When using Euler angles,
volume is computed with respect to the integration measure
\beq
dR = \frac{1}{8\pi^2} \sin \alpha \, d\alpha d\beta d\gamma,
\label{drdef}
\eeq
which when integrated over $0 \leq \alpha,\gamma \leq 2\pi$ and $0 \leq \beta \leq \pi$ gives a value of $1$.
Indeed, this result was obtained by construction by using the normalization of $8\pi^2$. The same volume
element will take on a different form when using the axis-angle parametrization, in analogy with the way
that the volume element in $\IR^3$ can be expressed in the equivalent forms
$dxdydz$ and $r^2\sin\theta dr d\phi d\theta$ in Cartesian and spherical coordinates, respectively.

Given any 3-parameter description of rotation,
the angular velocity of a rigid body can be obtained from a rotation matrix. Angular velocity in the
body-fixed and space-fixed reference frames can be written respectively as
$$ \bfomega_r = J_r(\qq) \dot{\qq} \askip \bfomega_l = J_l(\qq) \dot{\qq} $$
where $\qq$ is any parametrization (e.g., $\qq= [\alpha,\beta,\gamma]^T$ or $\qq = [\vartheta,\phi,\theta]^T$,
where $T$ denotes the transpose of a vector or matrix).

The Jacobian matrices $J_r(\qq)$ and $J_l(\qq)$ are computed from the parametrization $R(\qq)$ and
the definition of the $\vee$ operation in (\ref{vdefeq}) as

$$ {J}_l({\bf q}) =
\left[\left(\frac{\partial {R}}{\partial q_1} {R^T} \right)^{\vee}, \left(\frac{\partial {R}}{\partial q_2} {R^T} \right)^{\vee}, \left(\frac{\partial {R}}{\partial q_3} {R^T} \right)^{\vee} \right]. $$
and
$$ {J}_r({\bf q}) =
\left[\left({R^T} \frac{\partial {R}}{\partial q_1} \right)^{\vee}, \left({R^T} \frac{\partial {R}}{\partial q_2}\right)^{\vee},
\left({R^T} \frac{\partial {R}}{\partial q_3}\right)^{\vee} \right]. $$
This gives a hint as to why the subscripts $l$ and $r$ are used: if derivatives with respect to
parameters appear on the `right' of $R^T$, this is denoted with an $r$, and if they appear on the
`left' then a subscript $l$ is used.

Explicitly for the Euler angles,
\beq
{J}_l(\alpha,\beta,\gamma) =
\left[{\bf e}_3, {R_3}(\alpha) {\bf e}_1, {R_3}(\alpha){R_1}(\beta) {\bf e}_3 \right]
=
\left(\begin{array}{ccr}
0 & \cos \alpha & \sin \alpha \sin \beta \\
0 & \sin \alpha & -\cos \alpha \sin \beta \\
1 & 0 & \cos \beta
\end{array}\right)
\label{jleul}
\eeq
and
\beq
{J}_r = {R^T} J_{l} =
\left[{R_3}(-\gamma) {R_1}(-\beta)
{\bf e}_3, {R_3}(-\gamma) {\bf e}_1, {\bf e}_3 \right]
=  \left(\begin{array}{crc}
\sin \beta \sin \gamma & \cos \gamma & 0 \\
\sin \beta \cos \gamma & -\sin \gamma & 0 \\
\cos \beta & 0 & 1 \end{array}\right).
\label{jreul}
\eeq

Note that
$$ |J_l| = |J_r| = \sin \beta $$
gives the factor that appears in the volume element $dR$ in (\ref{drdef}). This is not a coincidence.
For any parametrization of $SO(3)$ of the form $R(\qq)$, the volume element can be expressed as
$$ dR = \frac{1}{8\pi^2} |J(\qq)| dq_1 dq_2 dq_3 $$
where $J(\qq)$ can be taken to be either $J_r(\qq)$ or $J_l(\qq)$. Though these matrices are not equal,
their determinants are.

Whereas the set of all rotations together with matrix multiplication forms a noncommutative ($R_1 R_2 \neq R_2 R_1$
in general) Lie group, the set of all angular velocity vectors $\bfomega_r$ and $\bfomega_l$ (or more precisely, their corresponding matrices, $\hat{\bfomega}_r$ and $\hat{\bfomega}_l$) together with the operations of addition
and scalar multiplication form a vector space. Furthermore, this vector space is endowed with an additional operation, the cross product $\bfomega_1 \times \bfomega_2$ (or equivalently the matrix commutator
$[\hat{\bfomega}_1, \hat{\bfomega}_2] = \hat{\bfomega}_1 \hat{\bfomega}_2 - \hat{\bfomega}_2 \hat{\bfomega}_1$).
This makes the set of all angular velocities a Lie algebra, which is denoted as $so(3)$ (as opposed to the
Lie group, $SO(3))$.

The Lie algebra $so(3)$ consists of skew-symmetric matrices of the form
\begin{equation}
{X} = \left(\begin{array}{ccc}
0 & -x_3 & x_2 \\
x_3 & 0 & -x_1 \\
-x_2 & x_1 & 0
\end{array} \right) = \sum_{i=1}^{3} x_i X_i.
\label{skew}
\end{equation}
The skew-symmetric matrices $\{X_i\}$ form a basis for the set of all such $3\times 3$ skew-symmetric matrices,
and the coefficients $\{x_i\}$ are all real.

Lie algebras and Lie groups are related in general by the exponential map. For matrix Lie groups
(which are the only kind of Lie groups that will be discussed here), the exponential map is the matrix
exponential function. In this specific case,
$$ \exp: so(3) \,\,\, \longrightarrow \,\,\, SO(3). $$

It is well known (see \cite{book} for derivation and references) that
\begin{equation}
R({\bf x}) = e^{X} = I + \frac{\sin \|{\bf x}\|}{\|{\bf x}\|} {X}
+ \frac{(1-\cos \|{\bf x}\|)}{\|{\bf x}\|^2} {X}^2
\label{matexp2}
\end{equation}
where $\|{\bf x}\| = (x_1^2 + x_2^2 + x_3^2)^{\half}$. Indeed, (\ref{matexp2}) is simply a variation on
(\ref{axisangle}) with $\xx = \vartheta {\bf n}$.

An interesting and useful fact is that except for a set of measure zero, all elements of $SO(3)$ can be captured
with the parameters within the open ball defined by $\|{\bf x}\| < \pi$, and the matrix logarithm of any group element
parameterized in this range is also well defined. It is convenient to know that the angle of the rotation, $\vartheta(R)$,
is related to the exponential parameters as
 $|\vartheta(R)| = \|{\bf x}\|$. Furthermore,
$$ \log(R) = \half \frac{\vartheta(R)}{\sin \vartheta(R)}
(R - R^T) $$
where
$$ \vartheta(R) = \cos^{-1}\left(\frac{{\rm trace}(R) -1}{2}\right). $$

Relatively simple analytical expressions have been derived for
the Jacobian $J_l$ and its inverse when rotations are parameterized
as in (\ref{matexp2}):
\begin{equation}
J_l({\bf x}) = I + \frac{1 - \cos \|{\bf x}\|}{\|{\bf x}\|^2}
{X} + \frac{\|{\bf x}\| - \sin \|{\bf x}\|}{\|{\bf x}\|^3} {X}^2.
\label{jlx928}
\end{equation}

The corresponding Jacobian $J_r$ is calculated
as \cite{book}
$$ J_r({\bf x}) = I - \frac{1 - \cos \|{\bf x}\|}{\|{\bf x}\|^2}
{X} + \frac{\|{\bf x}\| - \sin \|{\bf x}\|}{\|{\bf x}\|^3} {X}^2. $$

Note that
$$ J_l = J_{r}^{T} \askip J_l = R J_r. $$
The determinants are
$$ |\det(J_l)| = |\det(J_r)| = \frac{2(1-\cos\|{\bf x}\|)}{\|{\bf x}\|^2}. $$

\subsubsection{Example 2: The Euclidean Motion Group of the Plane}

The Euclidean motion group of the plane can be thought of as the set of all matrices of the form
\beq
g(x_1,x_2,\theta) = \left(\begin{array}{ccc}
\cos\theta & -\sin\theta & x \\
\sin\theta & \cos\theta & y \\
0 & 0 & 1 \end{array}\right)
\label{se2mainparam}
\eeq
together with the operation of matrix multiplication.

It is straightforward to verify that the form of these matrices is closed under multiplication and inversion,
and that $g(0,0,0) = \II$, and that it is therefore a group. This is often referred to as the special Euclidean
group, and is denoted as $SE(2)$. Like $SO(3)$, $SE(2)$ is three dimensional. However, unlike $SO(3)$,
$SE(2)$ is not compact. Nevertheless, it is possible to define a natural integration measure for $SE(2)$
as
$$ dg = dx dy d\theta. $$
And while $SE(2)$ does not have finite volume (and so there is no single natural normalization constant
such as $8\pi^2$ in the case of $SO(3)$), this integration measure nevertheless can be used to compute probabilities from probability densities.

Note that
$$ g(x ,y ,\theta) =  \exp(x X_1 + y X_2) \exp(\theta X_3) $$
where
$$ X_1 = \left(\begin{array}{ccc}
0 & 0 & 1 \\
0 & 0 & 0 \\
0 & 0 & 0 \end{array}\right); \hskip 0.2 true in
X_2 = \left(\begin{array}{ccc}
0 & 0 & 0 \\
0 & 0 & 1 \\
0 & 0 & 0 \end{array}\right); \hskip 0.2 true in
X_3 = \left(\begin{array}{ccc}
0 & -1 & 0 \\
1 & 0 & 0 \\
0 & 0 & 0 \end{array}\right). $$
These matrices form a basis for the Lie algebra, $se(2)$.
It is convenient to identify these with the natural basis for $\IR^3$ by
defining $(X_i)^{\vee} = {\bf e}_i$. In so doing, any element of $se(2)$ can be identified with
a vector in $\IR^3$.

The Jacobians for this parametrization
are then of the form
$$ J_l = \left[\begin{array}{ccc}
\left(\frac{\partial g}{\partial x} g^{-1}\right)^{\vee}, &
\left(\frac{\partial g}{\partial y} g^{-1}\right)^{\vee}, &
\left(\frac{\partial g}{\partial \theta} g^{-1}\right)^{\vee}\end{array}\right] =
\left(\begin{array}{ccc}
\cos \theta & \sin \theta & 0 \\
-\sin \theta & \cos \theta & 0 \\
0 & 0 & 1
\end{array}\right)
$$
and
$$ J_r = \left[\begin{array}{ccc}
\left(g^{-1} \frac{\partial g}{\partial x} \right)^{\vee}, &
\left(g^{-1} \frac{\partial g}{\partial y}  \right)^{\vee}, &
\left(g^{-1} \frac{\partial g}{\partial \theta}  \right)^{\vee}\end{array}\right] =
\left(\begin{array}{ccc}
1 & 0 & y \\
0 & 1 & -x \\
0 & 0 & 1
\end{array}\right). $$
Note that
$$ |{\rm det}(J_l)| = |{\rm det}(J_r)| = 1. $$
This parametrization is not unique, though it is probably the most well-known one.

As an alternative, consider the exponential parametrization $\exp: se(2) \rightarrow SE(2)$:
\beaq
g(x_1,x_2,x_3) &=& \exp(x_1 X_1 + x_2 X_2 + x_3 X_3)  \nonumber \\ &=& \exp \left(\begin{array}{ccc}
0 & -x_3 & x_1 \\
x_3 & 0 & x_2 \\
0 & 0 & 0 \end{array}\right) \nonumber \\ \nonumber \\
&=&
\left(\begin{array}{ccc}
\cos x_3 & -\sin x_3 &
[x_2(-1 + \cos x_3) + x_1\sin x_3]/x_3 \\
\sin x_3 & \cos x_3 &
[x_1(1 - \cos x_3) + x_2\sin x_3]/x_3
 \\
0 & 0 & 1 \end{array}\right).
\label{se2expparam}
\eeaq
Comparing this with (\ref{se2mainparam}) it is clear that $x_3 = \theta$, but $x \neq x_1$ and $y \neq x_2$.

The Jacobians in this exponential parametrization are
$$ J_r = \left(\begin{array}{ccc}
\frac{\sin x_3}{x_3} & \frac{\cos x_3 -1}{x_3} & 0 \\
\frac{1-\cos x_3}{x_3} & \frac{\sin x_3}{x_3} & 0 \\
\frac{x_3 x_1 - x_2 + x_2 \cos x_3 - x_1 \sin x_3}{x_3^2} &
\frac{x_1 + x_3 x_2 - x_1 \cos x_3 - x_2 \sin x_3}{x_3^2} & 1 \end{array} \right) $$

$$ J_l = \left(\begin{array}{ccc}
\frac{\sin x_3}{x_3} & \frac{1 - \cos x_3}{x_3} & 0 \\
\frac{\cos x_3 -1}{x_3} & \frac{\sin x_3}{x_3} & 0 \\
\frac{x_3 x_1 + x_2 - x_2 \cos x_3 - x_1 \sin x_3}{x_3^2} &
\frac{-x_1 + x_3 x_2 + x_1 \cos x_3 - x_2 \sin x_3}{x_3^2} & 1 \end{array} \right). $$
It follows that
$$ |{\rm det}(J_l)| = |{\rm det}(J_r)| =
\frac{2(1- \cos x_3)}{x_3^2}. $$

\subsubsection{Example 3: The Heisenberg Group}

The Heisenberg group, $H(1)$, is defined by elements of the form
\beq
g(\alpha, \beta, \gamma) = \ba{ccc}
1 & \alpha & \beta \\
0 & 1 & \gamma \\
0 & 0 & 1
\ea \htab \htab {\rm where} \htab \htab \alpha, \beta, \gamma \in \IR
\label{heissdef}
\eeq
and the operation of matrix multiplication. Therefore, the group law
can be viewed in terms of parameters as
$$ g(\alpha_1, \beta_1, \gamma_1) g(\alpha_2, \beta_2, \gamma_2) =
g(\alpha_1+\alpha_2, \beta_1+\beta_2 + \alpha_1\alpha_2, \gamma_1+\gamma_2). $$
The identity element is the identity matrix $g(0,0,0)$, and the
inverse of an arbitrary element $g(\alpha, \beta, \gamma)$ is
$$ g^{-1}(\alpha, \beta, \gamma) = g(-\alpha, \alpha \gamma - \beta, -\gamma). $$

Basis elements for the Lie algebra are
\beq
X_1 = \ba{ccc}
0 & 1 & 0 \\
0 & 0 & 0 \\
0 & 0 & 0
\ea ; \htab\htab
X_2 = \ba{ccc}
0 & 0 & 1 \\
0 & 0 & 0 \\
0 & 0 & 0
\ea ; \htab\htab
X_3 = \ba{ccc}
0 & 0 & 0 \\
0 & 0 & 1 \\
0 & 0 & 0
\ea .
\eeq

The Lie bracket, $[X_i,X_j] = X_i X_j - X_j X_i$, for these basis elements gives
$$ [X_1,X_2] = [X_2,X_3]=0 \htab\htab {\rm and} \htab\htab [X_1,X_3]=X_2. $$
If the inner product for the Lie algebra spanned by these basis
elements is defined as $(X,Y) = {\rm tr}(X Y^T)$, then this basis
is orthonormal: $(X_i,X_j) = \delta_{ij}$.

The group $H(1)$ is nilpotent because $(x_1 X_1 + x_2 X_2 + x_3 X_3)^{n} = 0$
for all $n \geq 3$. As a result, the matrix exponential is a polynomial in
the coordinates $\{x_i\}$:
\beq
\exp \ba{ccc}
0 & x_1 & x_2 \\
0 & 0 & x_3 \\
0 & 0 & 0
\ea = g(x_1, x_2 + \half{x_1 x_3}, x_3).
\label{expparheis}
\eeq
The parametrization in (\ref{heissdef}) can be viewed as the
following product of exponentials:
$$ g(\alpha, \beta, \gamma) = g(0, \beta, 0) g(0,0, \gamma) g(\alpha, 0,0) = \exp(\beta X_2) \exp(\gamma X_3) \exp(\alpha E_1). $$

The logarithm is obtained by solving for each $x_i$ as a function
of $\alpha, \beta, \gamma$. By inspection this is
$x_1 = \alpha$, $x_3 = \gamma$ and $x_2 = \beta - \alpha\gamma/2$.
Therefore,
$$ \log g(\alpha, \beta, \gamma) = \ba{ccc}
0 & \alpha &  \beta - \alpha\gamma/2 \\
0 & 0 & \gamma \\
0 & 0 & 0
\ea . $$

The Jacobian matrices for this group can be computed in either parametrization. In terms of $\alpha, \beta, \gamma$,
\beq
J_r(\alpha, \beta, \gamma) = \ba{ccr}
1 & 0 & 0 \\
0 & 1 & -\alpha \\
0 & 0 & 1
\ea \htab\htab {\rm and} \htab\htab
J_l(\alpha, \beta, \gamma) = \ba{rcc}
1 & 0 & 0 \\
-\gamma & 1 & 0 \\
0 & 0 & 1
\ea .
\label{leftheis}
\eeq
In terms of exponential coordinates,
\beq
J_r(\xx) = \ba{ccc}
1 & 0 & 0 \\
x_3/2 & 1 & -x_1/2 \\
0 & 0 & 1
\ea \htab\htab {\rm and} \htab\htab
J_{l}(\xx) = \ba{ccc}
1 & 0 & 0 \\
-x_3/2 & 1 & x_1/2 \\
0 & 0 & 1
\ea .
\label{expheis}
\eeq
In both parametrizations
$$ |{\rm det} J_r| = |{\rm det} J_l| = 1. $$

\subsubsection{Example 4: The Special Linear Group}

The group $SL(2,\IR)$ consists of all $2\times 2$ matrices with real
entries with determinant equal to unity. In other words, for
$a, b, c,d \in \IR$ elements of $SL(2,\IR)$ are of the form
$$ g = \ba{cc}
a & b \\
c & d \ea \wskip ad - bc = 1.$$
Subgroups of $SL(2,\IR)$ include matrices of the form
$$ g_1(x) = \exp \ba{cc}
x & 0 \\
0 & {-x} \ea  = \ba{cc}
e^{x} & 0 \\
0 & e^{-x} \ea ; $$
$$ g_2(y) = \exp \ba{cc}
0 & y \\
0 & 0 \ea  = \ba{cc}
1 & y \\
0 & 1 \ea ; $$
$$ g_3(\theta) = \exp \ba{cc}
0 & -\theta \\
\theta & 0 \ea  = \ba{cc}
\cos \theta & -\sin \theta \\
\sin \theta &  \cos \theta \ea . $$

A basis for the Lie algebra $sl(2,\IR)$ is
$$ X_1 = \left(\begin{array}{cc}
0 & -1 \\
1 & 0 \end{array}\right); \hskip 0.2 true in
X_2 = \left(\begin{array}{cc}
1 & 0 \\
0 & -1 \end{array}\right); \hskip 0.2 true in
X_3 = \left(\begin{array}{cc}
0 & 1 \\
1 & 0 \end{array}\right). $$
An inner product can be defined in which this basis is orthonormal.

It can be shown that any $g \in SL(2,\IR)$ can be
expressed as a product of $g_1(x)$, $g_2(y)$, and $g_3(\theta)$.
This is called an {\it Iwasawa decomposition} of $SL(2,\IR)$.

The above $g_i$ are not the only subgroups of $SL(2,\IR)$
For example, exponentiating matrices of the form $\xi \cdot (X_3 + 2 X_2)$
results in a subgroup of matrices of the form
\index{Iwasawa decomposition} \index{decomposition!Iwasawa}
$$ g(\xi) = \ba{cc}
\cosh \xi & \sinh \xi \\
\sinh \xi  & \cosh \xi \ea . $$

The {\it Iwasawa decomposition} allows
one to write an arbitrary $g \in SL(2,\IR)$
in the form \cite{sugiura}
$$ g = g_1(\theta) g_2(t) g_3(\xi) $$
where
$$ u_1(\theta) = \exp(\theta X_1) = \left(\begin{array}{cc}
\cos \theta & - \sin \theta \\
\sin \theta & \cos \theta \end{array}\right); $$
$$ u_2(t) = \exp(t X_2) = \left(\begin{array}{cc}
e^{t} & 0 \\
0 & e^{-t}
\end{array}\right); $$
$$ u_3(\xi) = \exp(\frac{\xi}{2}(X_3 - X_1)) = \left(\begin{array}{cc}
1 & \xi \\
0 & 1
\end{array}\right). $$

In this parametrization the right Jacobian is
$$ J_r(\theta,t,\xi) = \half \left(\begin{array}{ccc}
e^{-2t} + e^{2t}(1+ \xi^2) & -2 e^{2t} \xi &
e^{2t} - e^{-2t}(1+ e^{4t} \xi^{2}) \\
-2\xi & 2 & 2\xi \\
-1 & 0 & 1 \end{array} \right). $$

The left Jacobian is
$$ J_l(\theta,t,\xi) = \half \left(\begin{array}{ccc}
2 & 0 & 0 \\
0 & 2\cos 2\theta & 2\sin 2\theta \\
-e^{2t} & -e^{2t} \sin 2\theta & e^{2t} \cos
2\theta \end{array} \right). $$

It is easy to verify that
$$ |{\rm det}(J_r(\theta,t,\xi))| = |{\rm det}(J_l(\theta,t,\xi))| = \half e^{2t}.$$
Hence, $SL(2,\IR)$ is {\it unimodular} (which means the determinants
of the left and right Jacobians are the same). \index{unimodular Lie group}

\subsection{Generalizations}

Whereas several low-dimensional examples of Lie groups were presented to make the discussion concrete,
a vast variety of different kinds of Lie groups exist. For example, the same constraints that were used
to define $SO(3)$ relative to $\IR^{3\times 3}$ can be used to define $SO(n)$ from $\IR^{n\times n}$. The
result is a Lie group of dimension $n(n-1)/2$ and has a natural volume element $dR$.
Similarly, the Euclidean motion group generalizes as
all $(n+1)\times (n+1)$ matrices of the form
\beq
g = \left(\begin{array}{cc}
R & {\bf t} \\
{\bf 0}^T & 1 \end{array}\right) = \left(\begin{array}{cc}
\II & {\bf t} \\
{\bf 0}^T & 1 \end{array}\right) \left(\begin{array}{cc}
R & {\bf 0} \\
{\bf 0}^T & 1 \end{array}\right)
\label{sendefsscf}
\eeq
resulting in $SE(n)$ having dimension $n(n+1)/2$ and natural volume element $dg = dR d{\bf t}$
where ${\bf t} \in \IR^n$ and $d{\bf t} = dt_1 dt_2 \cdots dt_n$ is the natural integration measure
for $\IR^n$.
The following subsections briefly review the general theory of Lie groups that will be relevant
when defining information-theoretic inequalities.

\subsubsection{Exponential, Logarithm, and Vee Operation}

In general an n-dimensional real matrix Lie algebra is defined by a basis consisting of
real matrices  $\{X_i\}$ for $i=1,...,n$ that is closed under the matrix commutator.
That is, $[X_i, X_j] = \sum_{i=1}^{n} C_{ij}^{k} X_k$ for some real numbers $\{C_{ij}^{k}\}$,
which are called the structure constants of the Lie algebra.

In a neighborhood around the identity of the corresponding Lie group, the parametrization
\beq
g(x_1,...,x_n) = \exp X \wskip X = \sum_{i=1}^{n} x_i X_i
\label{genexppar}
\eeq
is always valid in a region around the identity in the corresponding Lie group.
And in fact, for the examples discussed, this parametrization
is good over almost the whole group, with the exception of a set of measure zero.

The logarithm map
$$ \log g(\xx) = X $$
(which is the inverse of the exponential) is valid
except on this set of measure zero. It will be convenient in the analysis to follow to identify
a vector $\xx \in \IR^n$ as
\beq
\xx = (\log \, g)^{\vee}
\wskip (X_i)^{\vee} = {\bf e}_i.
\label{veelsldcwfg}
\eeq
Here $\{{\bf e}_i\}$ is the natural basis for $\IR^n$.

In terms of quantities that have been defined in the examples, the adjoint matrices $Ad$ and $ad$ are
the following matrix-valued functions:
\beq
Ad(g) = J_l J_r^{-1} \askip ad(X) = \log Ad(e^{X}) .
\label{jkdkdkcs}
\eeq
The dimensions of these square matrices is the same as the dimension of the Lie group, which can be very different
than the dimensions of the matrices that are used to represent the elements of the group. The function
$\Delta(g) = {\rm det} Ad(g)$ is called the modular function of $G$. For a unimodular Lie group, $\Delta(g) = 1$.

\subsubsection{Integration and Differentiation on Unimodular Lie Groups}

Unimodular Lie groups are defined by the fact that their integration measures are invariant under shifts
and inversions. In any parametrization, this measure (or the corresponding volume element) can be expressed
as in the examples by first computing a left or right Jacobian matrix and then setting $dg = |J(\qq)|dq_1 dq_2 \cdots dq_n$ where $n$ is the dimension of the group. In the special case when $\qq = \xx$ is the
exponential coordinates, then \cite{helgason}
$$ \int_G f(g) dg = \int_{\cal G} f(e^X) {\rm det}\left(\frac{1-e^{-ad(X)}}{ad (X)} \right) d\xx $$
where $\xx = X^{\vee}$ and $d\xx = dx_1 dx_2 \cdots dx_n$. In the above expression it makes sense to write the division of one matrix
by another because the involved matrices commute. The symbol ${\cal G}$ is used to denote the Lie algebra
corresponding to $G$. In practice the integral is performed over a subset of ${\cal G}$, which is equivalent
to defining $f(e^X)$ to be zero over some portion of ${\cal G}$.

Let  $f(g)$ be a probability density function (or pdf for short) on a Lie group
$G$. Then
$$ \int_G f(g) dg = 1 \askip f(g) \geq 0. $$
It can be shown that unimodularity implies the following equalities for arbitrary $h \in G$, which generally do not all hold
simultaneously for measures on nonunimodular Lie groups:
\beq
\int_{G} f(g^{-1}) dg = \int_G f(h \circ g) dg = \int_G f(g \circ h) dg = \int_G f(g) dg.
\label{invlalrd}
\eeq

Many different kinds of unimodular Lie groups exist. For example, $SO(3)$ is compact and therefore has finite
volume; $SE(2)$ belongs to a class of Lie groups that are called solvable, $H(1)$ belongs to a class called
nilpotent; and $SL(2,\IR)$ belongs to a class called semisimple. Each of these classes of Lie groups has
been studied extensively. But for the purpose of this discussion, it is sufficient treat them all within the
larger class of unimodular Lie groups.

Given a function $f(g)$, the left and right Lie derivatives are defined with respect to
any basis element of the Lie algebra $X_i \in {\cal G}$ as
\begin{equation}
\tilde{X}_i^r f(g) = \left.\left(\frac{d}{dt}f(g \circ {\rm exp}(t
{X}_i))\right)\right|_{t=0}
\askip
\tilde{X}_i^l f(g) = \left.\left(\frac{d}{dt}f({\rm exp}(-t
{X}_i) \circ g)\right)\right|_{t=0}
. \label{direcd}
\end{equation}

The use of $l$ and $r$ mimicks the way that the subscripts were used in the Jacobians $J_l$ and $J_r$
in the sense that if ${\rm exp}(t{X}_i)$ appears on the left/right then the corresponding derivative
is given an $l/r$ designation. This notation, while not standard in the mathematics literature, is
useful in computations because when evaluating left/right Lie derivatives in coordinates $g=g(\qq)$, the
left/right Jacobians enter in the computation as \cite{book}
\begin{equation}
\tilde{\bf X}^r f = [J_r(\qq)]^{-T} \nabla_{\qq} f \askip \tilde{\bf X}^l f = -[J_l(\qq)]^{-T} \nabla_{\qq} f
\label{direcdgsc43}
\end{equation}
where $\tilde{\bf X}^r =[\tilde{X}_1^r, ..., \tilde{X}_n^r]^T$, $\tilde{\bf X}^l =[\tilde{X}_1^l, ..., \tilde{X}_n^l]^T$,
and $\nabla_{\qq} = [\partial/\partial q_1, ..., \partial /\partial q_n]^T$ is the gradient operator
treating $\qq$ like Cartesian coordinates.

\subsection{Probability Theory and Harmonic Analysis on Unimodular Lie Groups}

Given two probability density functions $f_1(g)$ and $f_2(g)$, their convolution is
\beq
(f_1 * f_2)(g) = \int_G f_1(h) f_2(h^{-1} \circ g) dh.
\label{convdef22dd}
\eeq
Here $h \in G$ is a dummy variable of integration. Convolution inherits associativity from
the group operation, but since in general $g_1 \circ g_2 \neq g_2 \circ g_1$,
$(f_1 * f_2)(g) \neq (f_2 * f_1)(g)$.

For a unimodular Lie group, the convolution integral of the form in (\ref{convdef22dd})
can be written in the following equivalent ways:
\begin{eqnarray}
(f_{1} * f_{2})(g) &=& \int_G f_{1}(z^{-1}) f_{2}(z \circ g) dz \nonumber \\
&=& \int_G f_{1}(g \circ k^{-1}) f_{2}(k) dk
\label{altcon}
\end{eqnarray}
where the substitutions $z = h^{-1}$ and $k = h^{-1} \circ g$ have been made,
and the invariance of integration under shifts and inversions in (\ref{invlalrd}) is used.

A powerful generalization of classical Fourier analysis exists. It is built on families
of unitary matrix-valued functions of group-valued argument that are parametrized by values
$\lambda$ drawn from a set $\hat{G}$ and satisfy the homomorphism
property:
\beq
U(g_1 \circ g_2,\lambda) = U(g_1,\lambda) U(g_2,\lambda).
\label{homoprop}
\eeq
Using $*$ to denote the Hermitian conjugate, it follows
that
$$ \II = U(e,\lambda) = U(g^{-1} \circ g,\lambda) = U(g^{-1},\lambda) U(g,\lambda), $$
and so
$$ U(g^{-1},\lambda) = (U(g,\lambda))^{-1} = U^*(g,\lambda). $$

In this generalized Fourier analysis (called noncommutative harmonic analysis)
each $U(g,\lambda)$ is constructed to be {\it irreducible} in the sense that it is not possible to simultaneously
block-diagonalize $U(g,\lambda)$ by the same similarity transformation for
all values of $g$ in the group. Such a matrix function $U(g,\lambda)$ is called an {\it irreducible
unitary representation}. Completeness of a set of
representations means that every (reducible) representation can be
decomposed into a direct sum of the representations in the set.

Once a complete set of IURs is known for a unimodular Lie group,
the Fourier transform of a function on that group can be defined as
$$ \hat{f}(\lambda) = \int_{G} f(g) U(g^{-1},\lambda) dg . $$
Here $\lambda$ (which can be thought of as frequency) indexes the complete set of all IURs.
An inversion formula can be used to recover the original function from all of the Fourier transforms as
\begin{equation}
f(g) = \int_{\hat{G}} {\rm
trace}[\hat{f}(\lambda) U(g,\lambda)] d(\lambda) . \label{four}
\end{equation}
The integration measure $d(\lambda)$ on the dual (frequency) space $\hat{G}$
is very different from one group to another.
In the case of a compact Lie group, $\hat{G}$ is discrete, and the resulting inversion formula
is a series, much like the classical Fourier series for $2\pi$-periodic functions.

A convolution theorem follows from (\ref{homoprop}) as
$$ \widehat{(f_1 * f_2)}(\lambda) = \hat{f}_2(\lambda) \hat{f}_1(\lambda) $$
and so does the Parseval/Plancherel formula:
\beq
\int_{G} |f(g)|^{2} dg = \int_{\hat{G}} ||\hat{f}(\lambda)||^2 d(\lambda) .
\label{plancherel}
\eeq
Here $|| \cdot ||$ is the Hilbert-Schmidt (Frobenius) norm, and
$d(\lambda)$ is the dimension of the matrix $U(g,\lambda)$.

A useful definition is
$$ u({X}_i,\lambda) = \frac{d}{dt} \left(
U({\rm exp}(t {X}_i),\lambda) \right)|_{t=0}.$$
Explicit expressions for $U(g,\lambda)$ and
$u(X_i,\lambda)$ using the exponential map and corresponding
parameterizations for the groups $SO(3)$, $SE(2)$ and $SE(3)$ are given in \cite{wooramrobotica, Gelfand}.

As a consequence of these definitions, it can be shown that the following operational properties result
\cite{book}:
$$ \widehat{{X}_{i}^{r} f} = u({X}_i,\lambda) \hat{f}(\lambda)
\askip
\widehat{{X}_{i}^{l} f} = -\hat{f}(\lambda) u({X}_i,\lambda) . $$

This is very useful in probability problems because a diffusion equation with drift of
the form
\begin{equation}
\frac{\partial \rho(g;t)}{\partial t} = - \sum_{i=1}^{d} h_i(t) \,
\tilde{X}_i^{r}  \rho(g;t) + \half  \sum_{i,j=1}^{d} D_{ij}
\tilde{X}_i^{r} \tilde{X}_j^{r} \, \rho(g;t) \label{mainmo}
\end{equation}
(where $D = [D_{ij}]$ is symmetric and positive semidefinite and given initial conditions $\rho(g;0)
=\delta(g)$) can be solved in the dual space $\hat{G}$, and then the inversion
formula can convert it back. Explicitly,
\beq
\rho(g;t) = \int_{\hat{G}}  {\rm
trace}[\exp(t {\cal B}(\lambda)) U(g,\lambda)] d(\lambda)
\label{Fsol}
\eeq
where
$$ {\cal B}(\lambda) = \half \sum_{k,l=1}^{n}
D_{lk} \, u({X}_l,\lambda) u({X}_k,\lambda)  - \sum_{l=1}^{n} h_l
\, u({X}_l,\lambda). $$
The solution to this sort of diffusion equation is important as a generalization of the concept of a Gaussian
distribution. It has been studied extensively in the case of $G = SE(3)$ in the context of polymer statistical
mechanics and robotic manipulators \cite{gcpoly1,gcpoly2,gcpoly3}.
As will be shown shortly, some of the classical information-theoretic inequalities that follow
from the Gaussian distribution can be computed using the above analysis.

\section{Properties of Entropy and Relative Entropy on Groups}
\label{infogroupsec}

As defined earlier, the entropy of a pdf on a unimodular Lie group is
$$ S(f) = -\int_{G} f(g) \log f(g) \, dg. $$
For example, the entropy of a Gaussian distribution with covariance $\Sigma$ is
\begin{equation}
S(\rho(g;t)) = \log\{(2\pi e)^{n/2} |\Sigma(t)|^{\half}\}
\label{gaussent}
\end{equation}
where $\log = \log_e$.

The Kullback-Leibler distance between the pdfs $f_1(g)$ and $f_2(g)$
 on a Lie group $G$ naturally generalizes from its form in $\IR^n$ 
as
\beq
D_{KL}(f_1 \| f_2) = \int_G f_1(g) \log \left(\frac{f_1(g)}{f_2(g)}\right) dg .
\label{kulldistgroup}
\eeq
As with the case of pdfs in $\IR^n$, $D_{KL}(f_1 \| f_2) \geq 0$
with equality when $D_{KL}(f \| f) =0$. And if $D_{KL}(f_1 \| f_2) = 0$
then $f_1(g) = f_2(g)$ at ``almost all'' values of $g \in G$
(or, in probability terminology ``$f_1(g) = f_2(g)$ almost surely''.
That is, they must be the same up to a set of measure zero.

Something that is not true in $\IR^n$ that holds for a compact Lie group is that the limiting distribution is the number one.
If $f_2(g) =1$ is the limiting
distribution, then $D_{KL}(f_1 \| 1) = -S(f_1)$.

\subsection{Convolutions Generally Increase Entropy}

\vskip 0.1 true in
\noindent
{\bf Theorem 3.1}: Given pdfs $f_1(g)$ and $f_2(g)$ on the unimodular Lie group $G$,
\beq
S(f_1 * f_2) \geq {\rm max} \{S(f_1), S(f_2)\}.
\label{entincr}
\eeq
\vskip 0.1 true in
\noindent
{\bf Proof:}
Denote the result of an $n$-fold convolution on $G$ as
$$ {f}_{1,n}(g) = (f_1* f_2 * f_3 * \cdots * f_n)(g). $$
Recall that a single pairwise convolution is computed as
$$ f_{i,i+1}(g) = (f_{i}* f_{i+1})(g) = \int_G f_{i}(h) f_{i+1}(h^{-1} \circ g) dh = \int_G f_{i}(g \circ k^{-1}) f_{i+1}(k) dk \neq (f_{i+1}*f_{i})(g) \,. $$
The $n$-fold convolution can be computed by performing a series of pairwise convolutions and stringing them together using the associative law. Convolution of functions on the group inherits associativity from the group law, which is reflected in the notation
$$ f_{i,i+2}(g) = (f_{i}* f_{i+1}* f_{i+2})(g) = (f_{i} * f_{i+1,i+2})(g) = (f_{i,i+1} * f_{i+2})(g) $$
where
$$ (f_{i} * f_{i+1,i+2})(g) =  (f_{i}*(f_{i+1}* f_{i+2}))(g) \askip
(f_{i,i+1} * f_{i+2})(g) = ((f_{i}*f_{i+1}) * f_{i+2})(g). $$

Johnson and Suhov \cite{johnson1, johnson2} proved the following
result for compact Lie groups:
\beq
D_{KL}({f}_{1,n} \| 1) - D_{KL}({f}_{1,n-1} \| 1)
= -\int_{G} D_{KL}({f}_{1,n-1} \| R(h) {f}_{1,n}) \, f_n(h) \, dh
\label{johnsonlemma}
\eeq
where $(R(h) f)(g) = f(g\circ h)$ is the right shift operator.
Since the integrand on the right side of (\ref{johnsonlemma}) is
nonnegative at all values of $h$ (and in fact, strictly positive
unless all $f_i(g)$ are delta functions), this indicates that
$$ D_{KL}({f}_{1,n} \| 1) \leq D_{KL}({f}_{1,n-1} \| 1) $$
with equality only holding in pathological cases. And so
iterated convolutions lead to
$$ \lim_{n\rightarrow \infty}  D_{KL}({f}_{1,n} \| 1) = 0
\tab \Longrightarrow \tab {f}_{1,n}(g) = 1 \tab a.s. $$

A noncompact group can not have $f(g)=1$ as a limiting
distribution, and so it does not make sense in this case
to use the notation $D_{KL}({f}_{1,n} \| 1)$. Nevertheless,
essentially the same proof that gives (\ref{johnsonlemma})
can be used in the more general
case of not-necessarily-compact unimodular Lie groups to show
that entropy must increase as a result of convolution.
This can be observed by first expanding out $S(f_{1,n})$ as:
\beaq
S(f_{1,n})  &=& -\int_G f_{1,n}(g) \log f_{1,n}(g) dg \\ \nonumber \\
 &=& -\int_G (f_{1,n-1}*f_n)(g) \log f_{1,n}(g) dg \\ \nonumber \\
&=& -\int_G \left[\int_G f_{1,n-1}(g \circ h^{-1}) f_{n}(h) dh\right] \log f_{1,n}(g) dg \label{linsev} \\  \nonumber \\
&=& -\int_G \int_G f_{1,n-1}(g \circ h^{-1}) f_{n}(h) \log f_{1,n}(g) \,dg \, dh \label{lineigh} \\  \nonumber \\
&=& -\int_G \int_G f_{1,n-1}(k) f_{n}(h) \log f_{1,n}(k \circ h) \,dk\, dh \, . \label{linnnin}
\label{johns292u21}
\eeaq
In going from (\ref{linsev}) to (\ref{lineigh}) all that was done
was to reverse the order of integration (i.e., using Fubini's Theorem)
and in going from (\ref{lineigh}) to (\ref{linnnin}) the change of
variables $k= g \circ h^{-1}$ is used together with the invariance
of integration under shifts.

Next, observe that
\bea
S(f_{1,n-1}) &=& - \int_G f_{1,n-1}(k) \log f_{1,n-1}(k) dk \\ \\
&=& - \left(\int_G f_{1,n-1}(k) \log f_{1,n-1}(k) dk\right)\left(\int_G f_{n}(h) dh\right) \\ \\
&=& - \int_G \left(\int_G f_{1,n-1}(k) \log f_{1,n-1}(k) dk\right) f_{n}(h) dh \, .
\eea
and so
\bea
S(f_{1,n}) - S(f_{1,n-1}) &=& \int_G\left(\int_G f_{1,n-1}(k) \left[\log f_{1,n-1}(k) - \log f_{1,n}(k \circ h) \right] dk\right) f_{n}(h) dh \\ \\
&=& \int_G\left(\int_G f_{1,n-1}(k) \log \left[\frac{f_{1,n-1}(k)}{f_{1,n}(k \circ h)}\right] dk\right) f_{n}(h) dh \\ \\
&=& \int_G  D_{KL}({f}_{1,n-1} \| R(h) {f}_{1,n}) \, f_{n}(h) \, dh \\
&\geq& 0.
\eea
Since no direct comparison between $f_{1,n}$ and the uniform distribution is made, Johnson and Suhov's proof of (\ref{johnsonlemma})
that has been adapted above yields
$$ S(f_{1,n-1} * f_n) \geq S(f_{1,n-1}). $$
Essentially the same proof can be used to show that
$$ S(f_1 * f_{2,n}) \geq S(f_{2,n}). $$
In other words, convolution in either order increases entropy.

\subsection{Entropy Inequalities from Jensen's Inequality}

Jensen's inequality is a fundamental tool that is often used in deriving information-theoretic inequalities, as
well as inequalities in the field of convex geometry.
In the context of Lie groups, Jensen's inequality can be written as
\beq
\Phi\left(\int_G \phi(g) \rho(g) dg\right) \leq \int_G \Phi(\phi(g)) \rho(g) dg
\label{jensenlie}
\eeq
where $\Phi:\IR_{\geq 0} \rightarrow \IR$ is a convex function on the half infinite line, $\rho(g)$ is
a pdf, and $\phi(g)$ is another nonnegative measurable function on $G$.

Two important examples of $\Phi(x)$ are $\Phi_1(x) = -\log x$ and $\Phi_2(x) = + x \log x$.
If $G$ is compact, any constant function on $G$ is measurable.
Letting $\phi(g) = 1$ and $\Phi(x) = \Phi_2(x)$ then gives $0 \leq -S(f)$ for a pdf $f(g)$. In contrast, for any unimodular Lie group, letting $\rho(g) = f(g)$, $\phi(g) = [f(g)]^{\alpha}$ and $\Phi(x) = \Phi_1(x)$ gives
\beq
-\log\left(\int_G [f(g)]^{1+\alpha} dg \right) \leq \alpha S(f).
\label{kdkguuced}
\eeq
This leads to the following theorem.
\vskip 0.1 true in
\noindent
{\bf Theorem 3.2}: Let $\|\hat{f}(\lambda)\|$ denote the Frobenius norm and
$\|\hat{f}(\lambda)\|_2$ denote the induced $2$-norm of the Fourier transform of $f(g)$ 
and define
\beq
D_2(f) = -\int_{\hat{G}} \log \|\hat{f}(\lambda)\|_2^2 d(\lambda) \,,\, {D}(f) = - \int_{\hat{G}} \log  \|\hat{f}(\lambda)\|^2 d(\lambda) \,,\, 
\tilde{D}(f) = - \log  \int_{\hat{G}} \|\hat{f}(\lambda)\|^2 d(\lambda).
\label{almostulf}
\eeq
Then
\beq
S(f) \geq \tilde{D}(f) \askip {D}(f) \leq {D}_2(f)
\label{hsjdkll3d0cddw}
\eeq
and
\beq
D_2(f_1 * f_2) \geq D_2(f_1) + D_2(f_2) \askip D(f_1 * f_2) \geq D(f_1) + D(f_2).
\label{hsjdkll3d0cddw2}
\eeq
Furthermore, denote the unit Heaviside step function on the real line as $u(x)$ and let
\beq 
B = \int_{\hat{G}} u\left( \|\hat{f}(\lambda)\|\right) d(\lambda) .
\hskip 0.1 true in {\rm Then} \hskip 0.1 true in \tilde{D}(f) + \log B \leq {D}(f)/B .
\label{dk432dlflfbb}
\eeq
For finite groups $B=1$ for functions that have full spectrum, 
and for bandlimited expansions on other groups $B$ is finite.
\vskip 0.1 true in
\noindent
{\bf Proof}: Substituting $\alpha =1$ into (\ref{kdkguuced}) and using the Plancherel formula (\ref{plancherel})
yields 
$$ S(f) \geq -\log\left(\int_G [f(g)]^{2} dg \right) =
-\log\left(\int_{\hat{G}} \|\hat{f}(\lambda)\|^2 d(\lambda)\right) =  \tilde{D}(f). $$
The fact that $-\log x$ is a decreasing function and  $\|A \|_2 \leq \|A \|$ for all
$A \in \mathbb{C}^{n\times n}$ gives the second inequality in (\ref{hsjdkll3d0cddw}).

The convolution theorem together with the facts that both norms are submultiplicative, $-\log(x)$ is 
a decreasing function, and
the log of the product is the sum of the logs gives
$$ D(f_1 * f_2) = - \int_{\hat{G}} \log \|\widehat{f_1 * f_2}(\lambda)\|^2 d(\lambda) =
- \int_{\hat{G}} \log \|\hat{f_1}(\lambda) \hat{f_2}(\lambda)\|^2 d(\lambda) \geq D(f_1) + D(f_2). $$
An identical calculation follows for $D_2$. 
The statement in (\ref{dk432dlflfbb}) follows from the Plancherel formula (\ref{plancherel}) and using Jensen's inequality (\ref{jensenlie}) in the dual space $\hat{G}$ rather than on $G$:
\beq
\Phi\left(\int_{\hat{G}} \|\hat{\phi}(\lambda)\| {\rho}(\lambda) d(\lambda) \right) \leq \int_G \Phi(\|\hat{\phi}(\lambda)\|) {\rho}(\lambda) d(\lambda) \wskip \int_{\hat{G}} {\rho}(\lambda) d(\lambda) = 1
\askip {\rho}(\lambda) \geq 0.
\label{jensenduallie}
\eeq
Recognizing that when $B$ is finite ${\rho}(\lambda) = u\left( \|\hat{f}(\lambda)\|\right)/B$ becomes a probability
measure on this dual space, it follows that 
\bea
\tilde{D}(f) &=& -\log\left(\int_{\hat{G}} \|\hat{f}(\lambda)\|^2 d(\lambda)\right) =
-\log\left(B \int_{\hat{G}} \|\hat{f}(\lambda)\|^2 {\rho}(\lambda) d(\lambda)\right) \\
&\leq& -\log B - \int_{\hat{G}} \log\left(\|\hat{f}(\lambda)\|^2\right) {\rho}(\lambda) d(\lambda)
= -\log B + {D}(f)/B.
\eea
This completes the proof.

Properties of dispersion measures similar to $D(f)$ and $D_2(f)$ were studied in \cite{grenander},
but no connections to entropy were provided previously. By definition, bandlimited expansions have $B$ finite.
On the other hand, it is a classical result that for a finite group, $\Gamma$, the Plancherel formula is 
(see, for example, \cite{book}):
$$ \sum_{\gamma \in \Gamma} |f(\gamma)|^2 = \frac{1}{|\Gamma|} \sum_{k=1}^{\alpha} d_k^2 \|\hat{f}_k\|^2 $$
where $\alpha$ is the number of conjugacy classes of $\Gamma$ and $d_k$ is the dimension of $\hat{f}_k$. And by Burnside's formula $\sum_{k=1}^{\alpha} d_k^2 = |\Gamma|$ it follows that $B=1$ when all $\|\hat{f}_k\| \neq 0$.

\subsection{The Entropy Produced by Convolution on a Finite Group is Bounded}

Let $\Gamma$ be a finite group with $|\Gamma|$ elements $\{g_1,..., g_{|\Gamma|}\}$, and
let $\rho^{\Gamma}(g_i) \geq 0$ with $\sum_{i =1}^{|\Gamma|} \rho^{\Gamma}(g_i) =1$ define a probability density/distribution on $\Gamma$.
In analogy with how convolution and entropy are defined on a Lie group, $G$, they can also be defined on a finite
group, $\Gamma$ by using the Dirac delta function for $G$, denoted here as $\delta(g)$.
If $\Gamma < G$ (i.e., if $\Gamma$ is a subgroup of $G$), then letting
$$ \rho^G(g) = \sum_{i =1}^{|\Gamma|} \rho^{\Gamma}(g_i) \delta(g_i^{-1} \circ g) = \sum_{\gamma \in \Gamma} \rho^{\Gamma}(\gamma) \delta(\gamma^{-1} \circ g) $$
can be used to define a pdf on $G$ that is equivalent to a pdf on $\Gamma$ in the sense that
if the convolution of two pdfs on $\Gamma$ is
\beq
(\rho_1^\Gamma * \rho_2^\Gamma)(g_i) = \sum_{j=1}^{|\Gamma|} \rho_1^\Gamma(g_j) \rho_2^\Gamma(g_{j}^{-1} \circ g_i) \label{discconveq}
\eeq
then
\beq
(\rho_1^G * \rho_2^G)(g) = \sum_{\gamma \in \Gamma} (\rho_1^\Gamma * \rho_2^\Gamma)(\gamma)
\delta(\gamma^{-1} \circ g).
\label{clk39kdd}
\eeq

Given a finite group, $\Gamma$, let
$$ S({\rho}) = - \sum_{i=1}^{|\Gamma|} \rho(g_i) \log \rho(g_i) = - \sum_{\gamma \in \Gamma} \rho(\gamma) \log \rho(\gamma). $$
Unlike the case of differential/continuous entropy on a Lie group, $0 \leq S({\rho})$.

The following theorem describes how the discrete entropy of pdfs on $\Gamma$ behaves under convolution.
Since only finite groups are addressed, the superscript $\Gamma$ on the discrete values $\rho(g_i)$ are
dropped.

\vskip 0.1 true in
\noindent
{\bf Theorem 3.3}: The entropy of the convolution of two pdfs on a finite group is greater than either
of the entropies of the convolved pdfs and is no greater than the
sum of their individual entropies
\begin{equation}
{\rm max} \{S(\rho_1), S(\rho_2)\} \leq S(\rho_{1} * \rho_{2}) \leq S(\rho_{1}) + S(\rho_{2}).
\label{convineq}
\end{equation}
\vskip 0.1 true in
\noindent
{\bf Proof:} The lower bound follows in the same way as the proof given for Theorem *.1 with summation
in place of integration.
The entropy of convolved distributions on a finite group
can be bounded from above in the following way.

Since the convolution sum contains products of all pairs, and each product is positive, it follows that
$$ \rho_1(g_k) \rho_2(g_{k}^{-1} \circ g_i) \leq (\rho_1 * \rho_2)(g_i) $$
for all $k \in \{1,...,|\Gamma|\}$. Therefore, since $\log$ is a strictly increasing function, it
follows that
$$  -S(\rho_{1} * \rho_{2}) \geq \sum_{i=1}^{|\Gamma|} \left(\sum_{j=1}^{|\Gamma|} \rho_1(g_j) \rho_2(g_{j}^{-1} \circ g_i)\right) \log \left(\rho_1(g_k) \rho_2(g_{k}^{-1} \circ g_i) \right). $$
Since this is true for all values of $k$, we can bring the $\log$ term inside of the
summation sign and choose $k=j$. Then multiplying by $-1$, and using the properties of the
log function, we get
$$ S(\rho_{1} * \rho_{2})
 \leq -  \sum_{i=1}^{|\Gamma|} \sum_{j=1}^{|\Gamma|} \rho_1(g_j) \rho_2(g_{j}^{-1} \circ g_i)
 \log \rho_1(g_j) - \sum_{i=1}^{|\Gamma|} \sum_{j=1}^{|\Gamma|} \rho_1(g_j) \rho_2(g_{j}^{-1} \circ g_i)
\log \rho_2(g_{j}^{-1} \circ g_i) . $$
Rearranging the order of summation signs gives
\begin{equation}
 S(\rho_{1} * \rho_{2}) \leq -   \sum_{j=1}^{|\Gamma|} \rho_1(g_j) \log \rho_1(g_j) \left(\sum_{i=1}^{|\Gamma|} \rho_2(g_{j}^{-1} \circ g_i) \right)
  -  \sum_{j=1}^{|\Gamma|} \rho_1(g_j) \left(\sum_{i=1}^{|\Gamma|} \rho_2(g_{j}^{-1} \circ g_i)
\log \rho_2(g_{j}^{-1} \circ g_i) \right)
.
\label{hjhj}
\end{equation}
But summation of a function over a group is invariant under shifts. That is,
$$ \sum_{i=1}^{|\Gamma|} F(g_{j}^{-1} \circ g_i) = \sum_{i=1}^{|\Gamma|} F(g_i)
\hskip 0.3 true in {\rm or} \hskip 0.3 true in \sum_{\gamma \in \Gamma} F(\gamma^{-1} \circ g) = \sum_{\gamma \in \Gamma} F(g) . $$
Hence, the terms in parenthesis in (\ref{hjhj}) can be written by replacing $g_{j}^{-1} \circ g_i$ with
$g_i$ gives (\ref{convineq}).

\subsection{Entropy and Decompositions}

Aside from the ability to sustain the concept of convolution, one of the fundamental ways that groups
resemble Euclidean space is the way in which they can be decomposed. In analogy with the way that
an integral over a vector-valued function with argument $\xx \in \IR^n$ can be decomposed into integrals
over each coordinate, integrals over Lie groups can also be decomposed in natural ways. This has
implications with regard to inequalities involving the entropy of pdfs on Lie groups. Analogous expressions
hold for finite groups, with volume replaced by the number of group elements.

\subsubsection{Direct Products}

Given the direct product of two groups, $G_1 \times G_2$, and a probability density $f(g_1,g_2)$ with
$$ \int_G \int_G f(g_1,g_2) dg_1 dg_2 = 1 $$
and the corresponding entropy is
$$ S_{12} = -\int_G \int_G f(g_1,g_2) \log f(g_1,g_2)  dg_1 dg_2. $$
\begin{equation}
In exact analogy with classical information theory, we can write 
S_{12} \leq S_1 + S_2
\label{marginal}
\end{equation}
where
$$ f_1(g_1) = \int_{G} f(g_1,g_2) dg_2 \askip f_2(g_2) = \int_{G} f(g_1,g_2) dg_1, $$
and
$$ S_i = -\int_{G} f_i(g_i) \log f_i(g_i) dg_i. $$
Equality in (\ref{marginal}) holds if and only if
$f(g_1,g_2) = f_1(g_1) f_2(g_2)$.

As in the case of pdfs on Euclidean space, (\ref{marginal}) follows from the fact that the Kullback-Leibler
divergence in (\ref{kulldistgroup}) has the property that
$D_{KL}(f \, \| \, f_1 f_2) \geq 0. $

\subsubsection{Coset Decompositions}

Given a subgroup $H \leq G$, and any element
$g\in G$, the {\it left coset} $gH$ is defined
as $ gH = \{g\circ h| h \in H\}. $
Similarly, the right coset $Hg$ is defined as
$ Hg = \{h\circ g| h \in H\}. $
In the special case when $g \in H$,
the corresponding left and right cosets are equal to $H$.
More generally for all $g\in G$,
$g \in gH$ and
$g_1 H = g_2 H$ if and only if $g_{2}^{-1} \circ g_1
\in H$. Likewise for right cosets
$H g_1 = H g_2 $ if and only if $g_1 \circ
g_{2}^{-1} \in H$.
Any group is divided into disjoint left (right) cosets,
and the statement ``$g_1$ and $g_2$ are
in the same left (right) coset'' is an
equivalence relation.

An important property of $gH$ and $Hg$
is that they have
the same number of elements as $H$. Since the
group is divided into disjoint cosets, each
with the same number of elements, it follows
that the number of cosets must divide without
remainder the number of elements in the group.
The set of all left(or right) cosets is called
the left(or right) {\it coset space}, and is denoted
as $G/H$ (or $H\backslash G$). For finite groups one writes
$ |G/H| = |H\backslash G| = |G|/|H|. $
This result is called {\it Lagrange's theorem}. Similar expressions can be written for Lie groups and Lie subgroups
after the appropriate concept of volume is introduced. We will use the following well-known fact \cite{helgason}:
\begin{equation}
\int_{G} f(g) d(g) =
\int_{G/H} \left(
\int_H f (g\circ h) d(h) \right) d(gH)
\label{quomeas2}
\end{equation}
where $g \in gH$ is taken to be the coset representative.
In the special case when $f(g)$ is a left-coset function (i.e., a function that is constant on left cosets),
(\ref{quomeas2}) reduces to
$$ \int_{G} f (g) d(g) = \int_{G/H} {F}(gH)  d(gH) $$
where it is assumed that $d(h)$ is normalized so
that ${\rm Vol}(H) = \int_{H} dh = 1$, and
$$ F(gH) = \int_H f (g\circ h) dh $$
is the value of the function $f(g)$ on each coset representative (which is the same as 
that which results from averaging over the coset $gH$).

\vskip 0.1 true in
\noindent
{\bf Theorem 3.4}: The entropy of a pdf on a unimodular Lie group is no greater than the sum of the marginal
entropies on a subgroup and the corresponding coset space:
\beq
S(f_G) \leq S(f_{G/H}) + S(f_H).
\label{skcnlrt543d}
\eeq

\vskip 0.1 true in
\noindent
{\bf Proof}:
For the moment it will be convenient to denote a function on $G$ as
$f_G(g)$ (rather than $f(g)$) and write
$$ f_G(g) = f_{G/H \times H}(g) = \tilde{f}_{G/H \times H}(gH, e). $$
That is, a function on $G$ evaluated at $g$ can be equally described as a function on a coset, together
with a rule for extracting a specific coset representative, which in this case is the identity.
This means that given $gH$, $g$ is recovered from $g \in gH$ as $g \circ e^{-1} = g$.
By enforcing the constraint on the definition of $\tilde{f}_{G/H \times H}$ that
$$ f_G(g \circ h) = \tilde{f}_{G/H \times H}(gH, h) \askip \tilde{f}_{G/H \times H}(H, h)
= \tilde{f}_{G/H \times H}(H, e), $$
then $g$ can be recovered from $g \circ h \in gH$ as $g \circ h \circ h^{-1} = g$.
Using this construction, we can define
$$ f_H(h) = \int_{G/H} f_G(g \circ h) d(gH) = \int_{G/H} \tilde{f}_{G/H \times H}(gH, h) d(gH) $$
and
$$ f_{G/H}(gH) = \int_{H} f_G(g \circ h) dh = \int_{H} \tilde{f}_{G/H \times H}(gH, h) dh. $$

For example, if $G=SE(n)$ is a Euclidean motion group and $H = SO(n)$ is the subgroup of pure rotations
in $n$-dimensional Euclidean space, then $G/H \cong \IR^n$, and we can write
$$ \int_{SE(n)} f(g) d(g) =
\int_{SE(n)/SO(n)} \left(
\int_{SO(n)} f (g\circ h) d(R) \right) d({\bf t})
$$

It follows from the classical information-inequality for the entropy of marginal distributions obtained by
letting $F(g) = -f(g)\log f(g)$ and using the nonnegativity of the Kullback-Leibler divergence
$$ D(f_G(g \circ h) \, \| \, f_{G/H} \cdot f_H(h)) \geq 0 $$
together with the shift-invariance of integrals on unimodular Lie groups that (\ref{skcnlrt543d}) holds.

\subsubsection{Double Coset Decompositions}

Let $H < G$ and $K < G$. Then for any $g \in G$,
the set
\begin{equation}
HgK = \{h\circ g\circ k|h\in H, k \in K\}
\label{doubledef}
\end{equation}
is called the {\it double coset} of $H$ and $K$,
and any $g^{'} \in HgK$ (including $g^{'}=g$)
is called a {\it representative} of the double coset.
Though a double coset representative often
can be described
with two or more different pairs $(h_1, k_1)$ and
$(h_2,k_2)$ so that $g^{'} = h_1 \circ g \circ k_1
= h_2 \circ g \circ k_2$, we only count $g^{'}$
once in $HgK$. Hence $|HgK| \leq |G|$, and in general
$|HgK| \neq |H|\cdot |K|$. In general, the set of
all double cosets of $H$ and $K$ is denoted
$H\backslash G/K$. Hence we have the hierarchy
$g \in HgK \in H\backslash G/K.$
It can be shown that membership in a double coset is an equivalence relation.
That is, $G$ is partitioned
into disjoint double cosets, and for
$H < G$ and $K<G$ either
$ H g_1 K \cap H g_2 K = \emptyset $
or $ Hg_1 K = H g_2 K$.

Another interesting thing to note (when certain
conditions are met) is the decomposition
of the integral of a function
on a group in terms of two subgroups
and a double coset space:
\beq
\int_{G} F(g) d(g) = \int_{K} \int_{K\backslash G/H}
\int_{H} F(k \circ g \circ h) d(h) d(KgH) d(k).
\label{doublecosetdecomp}
\eeq

A particular example of this is the integral over $SO(3)$,
which can be written in terms of Euler angles as

$$ \int_{SO(3)} dg = \int_{0}^{2\pi} \int_{0}^{\pi} \int_{0}^{2\pi} \frac{1}{8\pi^2} \sin \beta d\alpha d\beta d\gamma = $$
{\small
$$ \int_{SO(2)} \, \, \int_{SO(2)\backslash SO(3)/SO(2)} \, \,
\int_{SO(2)} \left(\frac{1}{2\pi} d\alpha\right) \left(\frac{1}{2} \sin \beta d\beta\right) \left(\frac{1}{2\pi} d\gamma\right). $$

\vskip 0.1 true in
\noindent
{\bf Theorem 3.5}: The entropy of a pdf on a group is no greater than the sum of marginal entropies over any two subgroups and the corresponding double-coset space:
\beq
S(f_G) \leq S(f_K) + S(f_{K\backslash G/H}) + S(f_H) .
\label{skcnlrt543jlvd}
\eeq

\vskip 0.1 true in
\noindent
{\bf Proof}:
Consistent with (\ref{doublecosetdecomp}) it is possible to decompose a function $f_G(g)$
as
$$ f_G(g) = \tilde{f}_{K \times K\backslash G/H \times H}(e, KgH, e) \wskip
 f_G(k \circ g \circ h) = \tilde{f}_{K \times K\backslash G/H \times H}(k, KgH, h)
. $$
If
$$ f_K(k) = \int_{K\backslash G/H} \int_H f_G(k \circ g \circ h) dh d(gH) $$
$$ f_H(h) = \int_K \int_{K\backslash G} f_G(k \circ g \circ h) d(Kg) dk $$
and
$$ f_{K\backslash G/H} = \int_K \int_H f_G(k \circ g \circ h) dh dk, $$
then letting $F(g) = -f(g)\log f(g)$ and using the nonnegativity of the Kullback-Leibler divergence
$$ D(f_G(k \circ g \circ h) \, \| \, f_K(k) \cdot f_{K\backslash G/H} \cdot f_H(h)) \geq 0 $$
together with the shift-invariance of integrals on unimodular Lie groups gives (\ref{skcnlrt543jlvd})

\subsubsection{Nested Coset Decompositions}

\vskip 0.1 true in
\noindent
{\bf Theorem 3.6}: The entropy of a pdf is no greater than the sum of entropies of its marginals over coset spaces
defined by nested subgroups:
\beq
S(f_G) \leq S(f_{G/K}) + S(f_{K/H}) + S(f_H).
\label{skcnlrt543dhh}
\eeq

\vskip 0.1 true in
\noindent
{\bf Proof}:
Given a subgroup $K$ of $H$, which is itself a subgroup of $G$ (that is, $H < K < G$),
it is possible to write \cite{helgason}
$$ \int_{G/H} F(gH) d(gH) = \int_{G/K} \left[\int_{K/H} F(g \circ k H) d(kH)\right] d(gK). $$
Therefore,
$$ \int_{G} F(g) dg = \int_{G/K} \int_{K/H} \int_H F(g \circ k \circ h) dh d(kH)  d(gK). $$
Again letting $F(g) = -f_G(g)\log f_G(g)$, it follows from the properties of Kullback-Leibler divergence
and the unimodularity of $G$ that if
$$ f_{G/K}(gK) = \int_{K/H} \int_H f(g \circ k \circ h) dh d(kH) $$
$$ f_{K/H}(kH) = \int_{G/K} \int_H f(g \circ k \circ h) dh  d(gK) $$
and
$$ f_H(h) = \int_{G/K} \int_{K/H} f(g \circ k \circ h) d(kH)  d(gK) $$
then (\ref{skcnlrt543dhh}) follows.

\subsubsection{Class Functions and Normal Subgroups}

In analogy with the way a coset is defined,
the conjugate of a subgroup $H$ for a given
$g \in G$ is defined as
$ gHg^{-1} = \{g\circ h \circ g^{-1}|h\in H\}.$
Recall that a subgroup $N \leq G$ is called {\it normal}
if and only if
$gNg^{-1} \subseteq N$ for all
$g \in G$. This is equivalent to the
conditions $g^{-1}Ng \subseteq N$, and so we also
write $g N g^{-1} =N$ and $gN = Ng$ for all
$g \in G$.

A function, $\chi(g)$, that is constant on each class has the property that
\beq
\chi(g) = \chi(h^{-1} \circ g \circ h) \hskip 0.1 true in {\rm or} \hskip 0.1 true in \chi(h \circ g) = \chi(g \circ h)
\label{class}
\eeq
for any $g,h \in G$.
Though convolution of functions on a noncommutative group is generally noncommutative, the special
nature of class functions means that
\bea
(f * \chi)(g) &=& \int_{G} f(h) \chi(h^{-1} \circ g) dh
= \int_{G} f(h) \chi(g \circ h^{-1}) dh \\
&=& \int_{G} \chi(k) f(k^{-1} \circ g)  dk = (\chi *f)(g).
\eea
where the change of variables $k = g \circ h^{-1}$ is used together with the unimodularity of $G$.

\subsection{When Inequivalent Convolutions Produce Equal Entropy}

In general $(\rho_1 * \rho_2)(g) \neq (\rho_2 * \rho_1)(g)$. Even so, it can be the case that
$ S(\rho_1 * \rho_2)(g) = S(\rho_2 * \rho_1)(g)$. This section addresses several special cases
when this equality holds.

Let $G$ denote a unimodular Lie group and for arbitrary $g, g_1 \in G$
define $\rho^{\vee}(g) = \rho(g^{-1})$,
$L_{g_1} \rho(g) = \rho(g_{1}^{-1} \circ g)$,
$R_{g_1} \rho(g) = \rho(g \circ g_{1})$,  $C_{g_1} \rho(g) = \rho(g_{1}^{-1} \circ g \circ g_{1})$.
Then if $\rho(g)$ is a pdf, it follows immediately
from (\ref{invlalrd}) that $\rho^{\vee}(g)$, $L_{g_1} \rho(g)$, $R_{g_1} \rho(g)$, and
$C_{g_1} \rho(g)$ are all pdfs.
A function for which $\rho^{\vee}(g) = \rho(g)$ is called symmetric, whereas
a function for which $C_{g_1} \rho(g) = \rho(g)$ for all $g_i \in G$ is a class function
(i.e., it is constant on conjugacy classes).

%\begin{theorem}
\vskip 0.1 true in
\noindent
{\bf Theorem 3.7}:
For arbitrary pdfs on a unimodular Lie group $G$ and arbitrary $g_1, g_2 \in G$,
$$ \rho_1 * \rho_2 \neq \rho^{\vee}_2 * \rho^{\vee}_1 \neq L_{g_1} \rho_1 * R_{g_2} \rho_2 \neq C_{g_1} \rho_1 * C_{g_1} \rho_2, $$
however, entropy satisfies the following equalities
\beq
S(\rho_1 * \rho_2) = S(\rho^{\vee}_2 * \rho^{\vee}_1) = S(L_{g_1} \rho_1 * R_{g_2} \rho_2) = S(C_{g_1} \rho_1 * C_{g_1} \rho_2).
\label{kskn3432sdrf}
\eeq
%\label{compentthmgen}
%\end{theorem}
\vskip 0.1 true in
\noindent
{\bf Proof:} Each equality is proven by changing variables and using the unimodularity property in (\ref{invlalrd}).
\bea (\rho^{\vee}_2 * \rho^{\vee}_1)(g) &=& \int_G \rho^{\vee}_2(h) \rho^{\vee}_1(h^{-1} \circ g) dh =
 \int_G \rho_2(h^{-1}) \rho_1(g^{-1} \circ  h) dh \\
&=& \int_G \rho_1(g^{-1} \circ  k^{-1}) \rho_2(k)  dk = (\rho_1 * \rho_2)(g^{-1}) = (\rho_1 * \rho_2)^{\vee}(g).
\eea
Let $F[\rho] = - \rho \log \rho$. Then due to (\ref{invlalrd}), the integral over $G$ of $F[\rho(g^{-1})]$
must be the same as $F[\rho(g)]$, proving the first equality in (\ref{kskn3432sdrf}). The second
equality follows from the fact that $(L_{g_1} \rho_1 * R_{g_2} \rho_2)(g) = (\rho_1 * \rho_2)(g_1 \circ g \circ g_2)$ and the integral of $F[\rho(g_1 \circ g \circ g_2)]$ must be the same as $F[\rho(g)]$, again due to (\ref{invlalrd}).
The final equality follows in a similar way from the fact that $(C_{g_1} \rho_1 * C_{g_1} \rho_2)(g) =  (\rho_1 * \rho_2)(g_1^{-1} \circ g \circ g_1)$.

Note that the equalities in (\ref{kskn3432sdrf}) can be combined. For example,
$$ S(\rho_1 * \rho_2) = S(L_{g_1} \rho^{\vee}_2 * R_{g_2} \rho^{\vee}_1) = S(C_{g_1} \rho^{\vee}_2 * C_{g_1} \rho^{\vee}_1). $$

\vskip 0.1 true in
\noindent
{\bf Theorem 3.8}:
%\begin{theorem}
The equality $S(\rho_1 * \rho_2) = S(\rho_2 * \rho_1)$ holds for pdfs $\rho_1(g)$ and $\rho_2(g)$ on a
unimodular Lie group $G$ in the following cases: (a) $\rho_i(g)$ for $i=1$ or $i=2$ is a class function; (b)
$\rho_i(g)$ for $i=1,2$ are both symmetric functions.
%\label{compentthm}
%\end{theorem}
\vskip 0.1 true in
\noindent
{\bf Proof:}
Statement (a) follows from the fact that if either $\rho_1$ or $\rho_2$ is a class function, then
convolutions commute.
Statement (b) follows from the first equality in (\ref{kskn3432sdrf}) and the definition of a symmetric function.

\vskip 0.1 true in
\noindent
%\begin{theorem}
{\bf Theorem 3.9}: Given class functions $\chi_1(g)$ and $\chi_2(g)$ that are pdfs, then for general
$g_1, g_2 \in G$,
$$ (\chi_1 * \chi_2)(g) \neq (L_{g_1} \chi_1 * L_{g_2} \chi_2)(g) \neq (R_{g_1} \chi_1 * R_{g_2} \chi_2)(g) \neq
(R_{g_1} \chi_1 * L_{g_2} \chi_2)(g) $$
and yet
\beq
S(\chi_1 * \chi_2) = S(L_{g_1} \chi_1 * L_{g_2} \chi_2) = S(R_{g_1} \chi_1 * R_{g_2} \chi_2) = S(R_{g_1} \chi_1 * L_{g_2} \chi_2).
\label{entclass30022}
\eeq
\vskip 0.1 true in
\noindent
{\bf Proof:}
\vskip 0.1 true in
Here the first and final equality will be proven. The middle one follows in the same way.
\bea
(L_{g_1} \chi_1 * L_{g_2} \chi_2)(g) &=& \int_{G} (L_{g_1} \chi_1)(h) * (L_{g_2} \chi_2)(h^{-1}\circ g) dh =
\int_{G} \chi_1(g_1^{-1} \circ h) \chi_2(g_2^{-1} \circ h^{-1}\circ g) dh \\
&=& \int_{G} \chi_1(k) \chi_2( g_2^{-1} \circ k^{-1}\circ g_1^{-1} \circ g) dk
= \int_{G} \chi_1(k) \chi_2(k^{-1}\circ g_1^{-1} \circ g \circ g_2^{-1}) dk \\
&=& (\chi_1 * \chi_2)(g_1^{-1} \circ g \circ g_2^{-1}).
\eea
Similarly,
\bea
(R_{g_1} \chi_1 * L_{g_2} \chi_2)(g) &=& \int_{G} (R_{g_1} \chi_1)(h) * (L_{g_2} \chi_2)(h^{-1}\circ g) dh =
\int_{G} \chi_1(h \circ g_1)  \chi_2(g_2^{-1} \circ h^{-1}\circ g) dh \\
&=& \int_{G} \chi_1(k) * \chi_2(g_2^{-1} \circ g_1 \circ k^{-1} \circ g) dk
= \int_{G} \chi_1(k) * \chi_2(k^{-1} \circ g  \circ  g_2^{-1} \circ g_1 ) dk
\\
&=& (\chi_1 * \chi_2)(g  \circ  g_2^{-1} \circ g_1).
\eea

\section{Fisher Information and Diffusions on Lie Groups} \label{fisherinfoliegr}

The natural extension of the Fisher information matrix for the case when
$f(g,\bftheta \,)$ is a parametric distribution on a Lie group is
\beq
F_{ij}(f, \bftheta \,) = \int_G \frac{1}{f} \frac{\partial f}{\partial \theta_i} \frac{\partial f}{\partial \theta_j} dg .
\label{fishmatgroupdef}
\eeq
In the case when $\bftheta$ parameterizes $G$ as
$g(\bftheta \,) = \exp(\sum_i \theta_i X_i)$ and
$f(g, \bftheta \,) = f(g \circ \exp(\sum_i \theta_i X_i))$,
then
$$ \left.\frac{\partial f}{\partial \theta_i}\right|_{\bftheta\,={\bf 0}} =
\xr_i f $$
and $F_{ij}(f, {\bf 0})$ becomes
\beq
F_{ij}^r(f) = \int_G \frac{1}{f} (\xr_i f)(\xr_j f) dg .
\label{fishmatgroupdef1}
\eeq
In a similar way, we can define
\beq
F_{ij}^l(f) = \int_G \frac{1}{f} (\xl_i f)(\xl_j f) dg .
\label{fishmatgroupdef2}
\eeq

\vskip 0.1 true in
\noindent
%\begin{theorem}
{\bf Theorem 4.1}:
The matrices (\ref{fishmatgroupdef1}) and (\ref{fishmatgroupdef1}) have the properties
\beq
F_{ij}^r(L(h)f) = F_{ij}^r(f) \askip F_{ij}^l(R(h)f) = F_{ij}^l(f) 
\label{rlshiftfish}
\eeq
and
\beq
F_{ij}^r(I(f)) = F_{ij}^l(f) \askip F_{ij}^l(I(f)) = F_{ij}^r(f)
\label{invfish}
\eeq
where $(L(h)f)(g) = f(h^{-1} \circ g)$, $(R(h)f)(g) = f(g \circ h)$, and $I(f)(g) = f(g^{-1})$.
\vskip 0.1 true in
\noindent
%\begin{theorem}
{\bf Proof}: The operators $\xl_i$ and $R(h)$ commute, and likewise $\xr_i$ and $L(h)$ commute.
This together with the invariance of integration under shifts proves (\ref{rlshiftfish}).
From the definitions of $\xl_i$ and $\xr_i$ in (\ref{direcd}), it follows that
$$ \tilde{X}_i^r (I(f))(g) = \left.\left(\frac{d}{dt}f([g \circ {\rm exp}(t
{X}_i)]^{-1})\right)\right|_{t=0} = \left.\left(\frac{d}{dt}f({\rm exp}(-t
{X}_i) \circ g^{-1})\right)\right|_{t=0} = (\tilde{X}_i^l f)(g^{-1}) . $$
Using the invariance of integration under shifts then gives (\ref{invfish}).
As a special case, when $f(g)$ is a symmetric function, the left and right Fisher information matrices
will be the same.

Note that the entries of Fisher matrices $F_{ij}^r(f)$ and $F_{ij}^l(f)$ implicitly depend on the choice of orthonormal Lie algebra basis $\{X_i\}$, and so it would be more descriptive to use the notation $F_{ij}^r(f,X)$ and $F_{ij}^l(f,X)$ .

If a different orthonormal basis $\{Y_i\}$ is used, such that
$X_i = \sum_{k} a_{ik} Y_k$, then the orthonormality of both
 $\{X_i\}$ and  $\{Y_i\}$ forces $A = [a_{ij}]$ to be an orthogonal
 matrix. Furthermore, the linearity of the Lie derivative,
$$ \tilde{X}^r f = \sum_{i} x_i \xr_i f \wskip X = \sum_i x_i X_i , $$
means that
$$
F_{ij}^{r}(f,X) = \int_G \frac{1}{f} \left(\sum_{k} a_{ik} \tilde{Y}^r_k f \right)\left(\sum_{l} a_{jl} \tilde{Y}^r_l f \right) dg = \sum_{k,l}
a_{ik} a_{jl} F_{kl}^{r}(f,Y).
$$
The same holds for $F_{ij}^{l}$. Summarizing these results in matrix form:
\beq
F^{r}(f,X) = A F^{r}(f,Y) A^T \askip F^{l}(f,X) = A F^{l}(f,Y) A^T
\wskip {\bf e}_i^T A {\bf e}_j = (X_i, Y_j).
\label{fishchobas}
\eeq
This means that the eigenvalues of the Fisher information matrix (and therefore its trace) 
are invariant under change of orthonormal basis.

\subsection{Fisher Information and Convolution on Groups}

The decrease of Fisher information as a result of convolution
can be studied in much the same way as for pdfs on Euclidean space.
Two approaches are taken here. First, a straightforward application
of the Cauchy-Bunyakovsky-Schwarz (CBS) inequality is used together
with the bi-invariance of the integral over a unimodular Lie group
to produce a bound on the Fisher information of the convolution
of two probability densities. Then, a tighter bound is obtained
using the concept of conditional expectation in the special case
when the pdfs commute under convolution.

\vskip 0.1 true in
\noindent
%\begin{theorem}
{\bf Theorem 4.2}: The following inequalities hold for the diagonal entries of the left and right Fisher
information matrices:
\beq
F_{ii}^r(f_1 * f_2) \leq {\rm min}\{F_{ii}^r(f_1),F_{ii}^r(f_2)\} \askip F_{ii}^l(f_1 * f_2) \leq {\rm min}\{F_{ii}^l(f_1),F_{ii}^l(f_2)\}  .
\label{fiineq92921}
\eeq
\vskip 0.1 true in
\noindent
%\begin{theorem}
{\bf Proof}:
The CBS inequality holds for groups:
$$ \left(\int_G a(g) b(g) dg\right)^2 \leq \int_G a^2(g) dg \int_G b^2(g) dg. $$
If $a(g) \geq 0$
 for all values of $g$, then it is possible to
 define $j(g) = [a(g)]^{\half}$ and $k(g)= [a(g)]^{\half} b(g)$, and
 since $j(g) k(g) = a(g) b(g)$,
\beaq
\left(\int_G a(g) b(g) dt\right)^2 &\leq& \left(\int_G j^2(g) dg\right) \left(\int_G k^2(t) dg \right) \nonumber\\ \nonumber\\
&=& \left(\int_G a(g) dg\right) \left(\int_G a(g) [b(g)]^2 dg \right) .
\label{csineq43432}
\eeaq
Using this version of the CBS inequality, and letting
$b(g) = \xr_i f_2(h^{-1}\circ g)/ [f_2(h^{-1}\circ g)]$ and $a(g) = f_1(h) f_2(h^{-1}\circ g) $,
essentially the same manipulations as in \cite{brown} can be used, with the roles of $f_1$ and $f_2$ interchanged due to the fact that
in general for convolution on a Lie group $(f_1*f_2)(g) \neq (f_2*f_2)(g)$:
\bea
F_{ii}^r(f_1 * f_2) &=&  \int_G \frac{\left(\int_G [\xr_i f_2(h^{-1}\circ g)/ f_2(h^{-1}\circ g)] \cdot [f_2(h^{-1}\circ g) f_1(h)] dh \right)^2}{(f_1*f_2)(g)} dg \\ \\
&\leq&  \int_G \frac{\left(\int_G [\xr_i f_2(h^{-1}\circ g)/ f_2(h^{-1}\circ g)]^2  [f_2(h^{-1}\circ g) f_1(h)] dh \right) \left(\int_G f_2(h^{-1}\circ g) f_1(h) dh\right)}{(f_1*f_2)(g)} dg \\ \\
&=& \int_G \left(\int_G \{[\xr_i f_2(h^{-1}\circ g)]^2/ f_2(h^{-1}\circ g) \} f_1(h) dh \right) dg \\ \\
&=& \int_G \left(\int_G \{[\xr_i f_2(h^{-1}\circ g)]^2/ f_2(h^{-1}\circ g) \}  dg \right) f_1(h) dh \\ \\
&=& F_{ii}^r(f_2) \int_G f_1(h) dh \\ \\
&=& F_{ii}^r(f_2)
\eea

Since for a unimodular Lie group it is possible to perform changes
of variables and inversion of the variable of integration without
affecting the value of an integral, the convolution can be written in the following equivalent ways,
\beaq
(f_1* f_2)(g) &=& \int_G f_1(h) f_2(h^{-1}\circ g) dh \\ \nonumber \\
&=& \int_G f_1(g \circ h^{-1}) f_2(h) dh \label{sknwekn3225} \\ \nonumber \\
&=& \int_G f_1(g \circ h) f_2(h^{-1}) dh \\ \nonumber \\
&=&  \int_G f_1(h^{-1}) f_2(h \circ g) dh
\eeaq
It then follows that using (\ref{sknwekn3225}) and the bi-invariance
of integration that (\ref{fiineq92921}) holds.

\subsubsection{A Tighter Bound Using Conditional Expectation for Commuting PDFs}

In this subsection a better inequality is derived.
\vskip 0.1 true in
\noindent
%\begin{theorem}
{\bf Theorem 4.3}: The following inequality holds for the right and left Fisher
information matrices:
\beq
{\rm tr}[F^r(\rho_1 * \rho_2) P] \leq {\rm tr}[F^r(\rho_i) P] \askip {\rm tr}[F^l(\rho_1 * \rho_2) P] \leq {\rm tr}[F^l(\rho_i) P]
\label{psjeneg}
\eeq
where $i=1,2$ and $P$ is an arbitrary symmetric positive definite matrix with the same dimensions as $F$.
\vskip 0.1 true in
\noindent
%\begin{theorem}
{\bf Proof}:
Let
$$ f_{12}(h,g) = \rho_1(h) \rho_2(h^{-1} \circ g). $$
Then
$$ f_1(h) = \int_G f_{12}(h,g) dg = \rho_1(h) \askip f_2(g) = \int_G f_{12}(h,g) dh = (\rho_1 * \rho_2)(g). $$
It follows that
$$ (\xr_i f_2)(g) = \int_G \rho_1(h) \xr_i \rho_2(h^{-1} \circ g) dh . $$
Then by the change of variables $k=h^{-1} \circ g$,
$$ (\xr_i f_2)(g) = \int_G \rho_1(g \circ k^{-1}) \xr_i \rho_2(k) dk . $$
This means that
\beq
\frac{(\xr_i f_2)(g)}{f_2(g)} = \int_G \frac{(\xr_i  \rho_2)(k)}{\rho_2(k)}
\frac{\rho_1(g \circ k^{-1}) \rho_2(k)}{f_2(g)} dk =
\left\langle \left. \frac{(\xr_i \rho_2)(k)}{\rho_2(k)} \right| g\right\rangle
\label{right20550}
\eeq
And therefore,
\bea
F_{ii}^r(f_2) &=& \left\langle \left(\frac{(\xr_i  \rho_2)(g)}{f_2(g)}\right)^2
\right\rangle = \left\langle \left\langle \left. \frac{(\xr_i \rho_2)(k)}{\rho_2(k)} \right| g\right\rangle^2 \right\rangle \\ \\
&\leq&  \left\langle \left\langle \left. \left(\frac{(\xr_i  \rho_2)(k)}{\rho_2(k)}\right)^2 \right| g\right\rangle \right\rangle
= \left\langle \left(\frac{(\xr_i \rho_2)(k)}{\rho_2(k)}\right)^2 \right\rangle \\ \\
&=& F_{ii}^r(\rho_2) .
\eea

An analogous argument using $ f_{12}(h,g) = \rho_1(g \circ h^{-1}) \rho_2(h) $ and $f_2(g) = (\rho_1* \rho_2)(g)$ shows that
\beq
\frac{(\xl_i f_2)(g)}{f_2(g)} =
\left\langle \left. \frac{(\xl_i \rho_1)(k)}{\rho_1(k)} \right| g\right\rangle
\label{left20550}
\eeq
and
$$ F_{ii}^l(f_2) \leq F_{ii}^l(\rho_1). $$

The above results can be written concisely by introducing an
arbitrary positive definite diagonal matrix $\Lambda$ as follows:
$$
{\rm tr}[F^r(\rho_1 * \rho_2) \Lambda] \leq {\rm tr}[F^r(\rho_2) \Lambda] \askip {\rm tr}[F^l(\rho_1 * \rho_2) \Lambda] \leq {\rm tr}[F^l(\rho_2) \Lambda] .
$$
If this is true in one basis, then using (\ref{fishchobas})
the more general statement in (\ref{psjeneg})
must follow in another basis where $P=P^T > 0$. Since the initial choice of basis is arbitrary, (\ref{psjeneg}) must hold in every basis for
an arbitrary positive definite matrix $P$. This completes the proof.

In some instances, even though the group is not commutative,
the functions $\rho_1$ and $\rho_2$ will commute. For example,
if $\rho(g \circ h) = \rho(h \circ g)$ for all $h,g \in G$, then
$(\rho * \rho_i)(g) = (\rho_i * \rho)(g)$ for any reasonable choice of
$\rho_i(g)$. Or if $\rho_2 = \rho_1 *\rho_1 * \cdots \rho_1$ it will clearly be the case that $\rho_1* \rho_2 = \rho_2 * \rho_1$.
If, for whatever reason, $\rho_1* \rho_2 = \rho_2 * \rho_1$ then
(\ref{psjeneg}) can be rewritten in the following form:
\beaq
{\rm tr}[F^r(\rho_1 * \rho_2) P] &\leq& {\rm min}\{{\rm tr}[F^r(\rho_1) P],{\rm tr}[F^r(\rho_2) P]\} \nonumber \\
&{\rm and}& \label{psjenegcom} \\
 {\rm tr}[F^l(\rho_1 * \rho_2) P] &\leq& {\rm min}\{{\rm tr}[F^l(\rho_1) P], {\rm tr}[F^l(\rho_2) P]\} \nonumber
\eeaq

\vskip 0.1 true in
\noindent
%\begin{theorem}
{\bf Theorem 4.4}: When $\rho_1* \rho_2 = \rho_2 * \rho_1$ the following equality holds
\beq
\frac{1}{{\rm tr}[F^r(\rho_1*\rho_2) P]} \leq
\frac{1}{{\rm tr}[F^r(\rho_1) P]} + \frac{1}{{\rm tr}[F^r(\rho_2) P]}
\tab {\rm for} \tab {\rm any} \tab P = P^T > 0,
\label{fishrffftight}
\eeq
and likewise for $F^l$.
\vskip 0.1 true in
\noindent
%\begin{theorem}
{\bf Proof}: Returning to (\ref{right20550}) and (\ref{left20550}), in the
case when $\rho_1* \rho_2 = \rho_2 * \rho_1$ it is possible to
write
\beq
\frac{(\xr_i f_2)(g)}{f_2(g)} =
\left\langle \left. \frac{(\xr_i \rho_2)(k)}{\rho_2(k)} \right| g\right\rangle =
\left\langle \left. \frac{(\xr_i \rho_1)(k)}{\rho_1(k)} \right| g\right\rangle
\label{kwbbo4442}
\eeq
and
$$
\frac{(\xl_i f_2)(g)}{f_2(g)} =
\left\langle \left. \frac{(\xl_i \rho_1)(k)}{\rho_1(k)} \right| g\right\rangle = \left\langle \left. \frac{(\xl_i \rho_2)(k')}{\rho_2(k')} \right| g\right\rangle. $$

Since the following calculation works the same way for both the
`l' and `r' cases, consider only the `r' case for now.
Multiplying the first equality in (\ref{kwbbo4442}) by $1-\beta$ and the second by $\beta$ and adding together\footnote{The names of the dummy variables $k$ and $k'$ are unimportant. However, at this stage it is important that the names be different in order to emphasize their statistical independence.}:
\bea
\frac{(\xr_i f_2)(g)}{f_2(g)}
 &=& \beta \left\langle \left. \frac{(\xr_i \rho_1)(k)}{\rho_1(k)} \right| g\right\rangle + (1-\beta)
 \left\langle \left. \frac{(\xr_i \rho_2)(k')}{\rho_2(k')} \right| g\right\rangle
  \\ \\
&=&  \left\langle \left. \beta \frac{(\xr_i \rho_1)(k)}{\rho_1(k)} + (1-\beta) \frac{(\xr_i \rho_2)(k')}{\rho_2(k')} \right| g \right\rangle
\eea
for arbitrary $\beta \in [0,1]$.

Now squaring both sides and taking the (unconditional) expectation,
and using Jensen's inequality yields:
\beaq
\left\langle
\left( \frac{(\xr_i f_2)(g)}{f_2(g)} \right)^2
\right\rangle
&=&
\left\langle
\left\langle \left. \beta \frac{(\xr_i \rho_1)(k)}{\rho_1(k)} + (1-\beta) \frac{(\xr_i \rho_2)(k')}{\rho_2(k')} \right| g \right\rangle^2
\right\rangle \nonumber \\ \nonumber \\
&\leq&
\left\langle
\left( \beta \frac{(\xr_i \rho_1)(k)}{\rho_1(k)} + (1-\beta) \frac{(\xr_i \rho_2)(k')}{\rho_2(k')} \right)^2
\right\rangle \nonumber \\  \nonumber \\
&=&
\beta^2 \, \left\langle \left(\frac{(\xr_i \rho_1)(k)}{\rho_1(k)} \right)^2 \right\rangle  +
(1-\beta)^2  \left\langle \left(\frac{(\xr_i \rho_2)(k')}{\rho_2(k')}
\right)^2
\right\rangle
\label{johndleleq929}
\eeaq

This statement simply says
\beq
F_{ii}^r(\rho_1*\rho_2) \leq \beta^2 \, F_{ii}^r(\rho_1) +
(1-\beta)^2 \, F_{ii}^r(\rho_2).
\label{kenebi44o4}
\eeq
The value of $\beta \in [0,1]$ that gives the tightest bound is
$$ \beta = \frac{F_{ii}^r(\rho_2)}{F_{ii}^r(\rho_1) +  F_{ii}^r(\rho_2)}, $$
resulting in the inequality
\beq
\frac{1}{F_{ii}^r(\rho_1*\rho_2)} \leq \frac{1}{F_{ii}^r(\rho_1)} +
\frac{1}{F_{ii}^r(\rho_2)} .
\label{fishineqtightii}
\eeq
Alternatively, if before computing the optimal $\beta$ we first
multiply both sides of (\ref{kenebi44o4}) by $\lambda_i$ and sum
over $i$, the result will be
$$ {\rm tr}[F^r(\rho_1*\rho_2) \Lambda] \leq \beta^2 \, {\rm tr}[F^r(\rho_1) \Lambda] +
(1-\beta)^2 \, {\rm tr}[F^r(\rho_2) \Lambda]. $$
Again, since the basis is arbitrary, $\Lambda$ can be replaced with $P$. Then the optimal value of $\beta$ will give (\ref{fishrffftight}).

\subsubsection{A Special Case: $SO(3)$}

Consider the group of $3\times 3$ orthogonal matrices with determinant
$+1$. Let ${\bf \xr}= [\xr_1, \xr_2, \xr_3]^T$ and ${\bf \xl}= [\xl_1, \xl_2, \xl_3]^T$. These two gradient vectors are related to each other
by an adjoint matrix, which for this group is a rotation matrix
\cite{book}. Therefore, in the case when $G=SO(3)$,
$$ \|{\bf \xr} f\|^2 = \|{\bf \xl} f\|^2 \tab \Longrightarrow \tab
{\rm tr}[F^r(f)] = {\rm tr}[F^l(f)] $$
Therefore, the inequalities in (\ref{psjenegcom})
will hold for pdfs on $SO(3)$ regardless
of whether or not the functions commute under convolution, but restricted to the condition $P=I$.

\section{Generalizing the de Bruijn Identity to Lie Groups} \label{deBruijngr}

This section generalizes the de Bruijn identity, in which entropy rates are related to
Fisher information.
\vskip 0.1 true in
\noindent
{\bf Theorem 5.1}:
%\begin{theorem}
% For any pdf $\rho(g)$, the entropy is defined as
% $$ S(\rho) = -\int_G \rho(g) \log \rho(g) dg. $$
Let $f_{D, {\bf h},t}(g) = f(g,t;D,{\bf h})$ denote the solution to the diffusion
equation (\ref{mainmo}) with constant ${\bf h}$ subject to the initial condition $f(g,0;D,{\bf h}) = \delta(g)$.
Then for any well-behaved pdf $\alpha(g)$,
\beq
\frac{d}{dt} S(\alpha* f_{D, {\bf h},t}) = \half {\rm tr}[D F^r(\alpha * f_{D, {\bf h},t})].
\label{debru323ii205}
\eeq
\vskip 0.1 true in
\noindent
{\bf Proof}:
It is easy to see that the solution of the diffusion equation
\beq
\frac{\partial \rho}{\partial t} = \half \sum_{i,j=1}^{n} D_{ij} \xr_i \xr_j \rho
- \sum_{k=1}^{n} h_k \xr_k \rho
\label{skk3k2s}
\eeq
subject to the initial conditions $\rho(g,0) = \alpha(g)$
 is simply $\rho(g,t) = (\alpha * f_{D,{\bf h},t)}(g)$. This follows
 because all derivatives ``pass through'' the convolution integral
 for $\rho(g,t)$ and act on $f_{D,{\bf h},t}(g)$.

 Taking the time derivative of $S(\rho(g,t))$ we get
\beq
\frac{d}{dt} S(\rho) = -\frac{d}{dt} \int_G \rho(g,t) \log \rho(g,t) dg = -\int_G \left\{
\frac{\partial \rho}{\partial t} \log \rho + \frac{\partial \rho}{\partial t} \right\} dg.
\label{skk322499}
\eeq
Using (\ref{skk3k2s}), the partial with respect to time can be replaced with Lie derivatives.
But
$$ \int_{G} \xr_k \rho dg = \int_{G} \xr_i \xr_j \rho dg = 0, $$
so the second term on the right side of (\ref{skk322499})
completely disappears. Using the integration-by-parts formula\footnote{There are no
surface terms because, like the circle and real line, each coordinate in the integral either wraps around
or goes to infinity.}
$$ \int_{G} f_1 \xr_k f_2 dg = - \int_{G} f_2 \xr_k f_1 dg, $$
with $f_1 = \log \rho$ and $f_2 = \rho$ then gives
\bea
 \frac{d}{dt} S(\alpha * f_{D,{\bf h},t}) &=& \half \sum_{i,j=1}^{n} D_{ij} \int_G  \frac{1}{\alpha * f_{D,{\bf h},t}} \xr_j (\alpha * f_{D,{\bf h},t}) \xr_i (\alpha * f_{D,{\bf h},t})\, dg \\
&=& \half \sum_{i,j=1}^{n} D_{ij} F^r_{ij}(\alpha * f_{D,{\bf h},t}) \\
&=& \half {\rm tr}\left[D F^r(\alpha * f_{D,{\bf h},t})\right].
\eea

The implication of this is that
$$ S(\alpha * f_{D,{\bf h},t_2}) - S(\alpha * f_{D,{\bf h},t_1}) = \half \int_{t_1}^{t_2} {\rm tr}\left[D F^r(\alpha * f_{D,{\bf h},t})\right] dt $$

% If $\alpha(g) = \delta(g)$ and $0< t_1, t_2 << 1$ can the Fisher
% information matrix be computed in closed form ?

\section{Information-Theoretic Inequalities from Log-Sobolev Inequalities} \label{logsoblie}

In this section information-theoretic identities are derived from Log-Sobolev inequalities.
Subsection \ref{lsrev} provides a brief review of Log-Sobolev inequalities. Subsection \ref{newlssec}
then uses these to write information-theoretic inequalities.

\subsection{Log-Sobolev Inequalities in $\IR^n$ and on Lie Groups} \label{lsrev}

The log-Sobolev inequality can be stated as
\cite{beckner1, beckner2, lieb}:
\beq
 \int_{\IR^n} |\psi(\xx)|^2 \log |\psi(\xx)|^2 d\xx \leq \frac{n}{2} \log \left[\frac{2}{\pi e n} \int_{\IR^n}
\|\nabla \psi\|^2 d\xx \right] 
\label{logsoboineq}
\eeq
where
$$ \nabla \psi = \left[\frac{\partial \psi}{\partial x_1}, ..., \frac{\partial \psi}{\partial x_n}\right]^T \askip \int_{\IR^n} |\psi(\xx)|^2 d\xx = 1. $$
Here $\log = \log_e$.
Actually, there is a whole family of log-Sobolev inequalities, and (\ref{logsoboineq}) represents the
tightest of these. The original form of the log-Sobolev inequality as introduced by Gross in \cite{gross1} is
\beq
\half \int_{\IR^n} |\phi(\xx)|^2 \log|\phi(\xx)|^2 \rho(\xx) d\xx \leq \int_{\IR^n} \|\nabla \phi(\xx)\|^2 \rho(\xx) d\xx
+ \|\phi\|_{L^2(\IR^n,\rho)}^2 \log \|\phi\|_{L^2(\IR^n,\rho)}^2
\label{logsoboineq2}
\eeq
where
$$ \|\phi\|_{L^2(\IR^n,\rho)}^2 = \int_{\IR^n} |\phi(\xx)|^2 \rho(\xx) d\xx . $$
Here $\rho(\xx) = \rho(\xx,0) = (2\pi)^{-n/2} \exp(-\|\xx\|^2/2)$ is the solution to the heat equation
on $\IR^n$ evaluated at $t=1$.

Several different variations exist. For example, by rescaling, it is possible to rewrite (\ref{logsoboineq2}) with
$\rho(\xx,t)$ in place of $\rho(\xx)$ by introducing a multiplicative factor of $t$ in the first term on the
right hand side of the equation. Or, by letting $\phi(\xx) = \rho^{-\half}(\xx) \psi(\xx/a)$ for some scaling
factor $a>0$, substituting into (\ref{logsoboineq2}), and integrating by parts then gives \cite{lieb}
$$ \int_{\IR^n} |\psi(\xx)|^2 \log\frac{|\psi(\xx)|^2}{\|\psi\|_2^2}  d\xx + n(1+\log a)\|\psi\|_2^2 \leq
\frac{a^2}{\pi} \int_{\IR^n} \|\nabla \psi(\xx)\|^2 d\xx $$
where
$$ \|\psi\|_2^2 = \int_{\IR^n} |\psi(\xx)|^2 d\xx \askip \|\nabla \psi(\xx)\|^2 = \nabla \psi(\xx) \cdot \nabla \psi(\xx). $$
This, together with an optimization over $a$ gives (\ref{logsoboineq}).

Gross subsequently extended (\ref{logsoboineq2}) to Lie groups \cite{gross2} as
\beq
\int_G \left\{|\phi(g)|^2 \log |\phi(g)|\right\} \rho(g,t) dg \leq c_G(t)
\int_G \|({\bf \tilde{X}}^r \phi)(g)\|^2 \rho(g,t) dg + \|\phi\|_{L^2(G,\rho_t)}^2 \log \|\phi\|_{L^2(G,\rho_t)}^2
\label{logsobolieineq2}
\eeq
where $\rho(g,t)$ is the solution to the diffusion equation in (\ref{skk3k2s}) with $h_i=0$, $D_{ij} = \delta_{ij}$, initial condition $\rho(g,0) = \delta(g)$, and 
$$ {\bf \tilde{X}}^r \phi = [\tilde{X}_1^r \phi, ..., \tilde{X}_n^r \phi]^T \askip
\|\phi\|_{L^2(G,\rho_t)}^2 = \int_G |\phi(g)|^2 \rho(g,t) dg . $$
In (\ref{logsobolieineq2}) the scalar function $c_G(t)$ depends on the particular group. For $G = (\IR^n,+)$
we have $c_{\IR^n}(t) =t$, and likewise $c_{SO(n)}(t) =t$.

In analogy with the way that (\ref{logsoboineq}) evolved from (\ref{logsoboineq2}), a descendent of (\ref{logsobolieineq2}) for noncompact unimodular Lie groups is \cite{bakry,beckner1,beckner2}
\beq
 \int_{G} |\psi(g)|^2 \log |\psi(g)|^2 dg \leq \frac{n}{2} \log \left[\frac{2 C_G}{\pi e n} \int_{G}
\|{\bf \tilde{X}} \psi\|^2 dg \right] 
\label{logsobolieineq3}
\eeq
The only difference is that, to the author's knowledge, the sharp factor $C_G$ in this expression is not known
for most Lie groups.
The information-theoretic interpretation of these inequalities is provided in the following subsection.

\subsection{Information-Theoretic Inequalities} \label{newlssec}

For our purposes the form in (\ref{logsoboineq}) will be most useful. It is interesting to note
in passing that Beckner has extended this inequality to the case where the domain, rather than being $\IR^n$,
is the hyperbolic space $\mathbb{H}^2 \cong SL(2,\IR)/SO(2)$ and the Heisenberg groups $H(n)$, including $H(1)$ \cite{beckner1,beckner2}.
Our goal here is to provide an information-theoretic interpretation of the inequalities from the previous section.

\vskip 0.1 true in
\noindent
{\bf Theorem 6.1}: Entropy powers and Fisher information are related as 
\beq
[N(f)]^{-1} \leq \frac{1}{n} {\rm tr}(F) \wskip N(f)=  \frac{C_G}{2\pi e} \exp \left[\frac{2}{n} S(f)\right].
\label{logsobfingr}
\eeq

\vskip 0.1 true in
\noindent
{\bf Proof}: We begin by proving (\ref{logsobfingr}) for $G = (\IR^n,+)$.
Making the simple substitution $f(\xx) = |\psi(\xx)|^2$
into (\ref{logsoboineq}) and requiring that $f(\xx)$ be a pdf gives
$$ \int_{\IR^n} f(\xx) \log f(\xx) d\xx \leq \frac{n}{2} \log \left[\frac{1}{2 \pi e n} \int_{\IR^n} \frac{1}{f}
\|\nabla f \|^2 d\xx \right]. $$
or
\beq
-S(f) \leq \frac{n}{2} \log \frac{{\rm tr}(F)}{2 \pi e n} \hskip 0.1 true in \Longrightarrow \hskip 0.1 true in
\exp\left[-\frac{2}{n} S(f)\right] \leq  \frac{{\rm tr}(F)}{2 \pi e n} 
\hskip 0.1 true in \Longrightarrow \hskip 0.1 true in  [N(f)]^{-1} \leq \frac{1}{n} {\rm tr}(F) .
\label{logsobofish}
\eeq
Here $S(f)$ is the Boltzmann-Shannon entropy of $f$ and $F$ is the Fisher information matrix.
As is customary in information theory, the entropy power can be defined as $N(f)$ in (\ref{logsobfingr}) 
with $C_G = 1$. Then the log-Sobolev inequality in the form in (\ref{logsobofish}) is written as 
(\ref{logsobfingr}).

For the more general case, starting with (\ref{logsobolieineq3}) and letting $f(g) = |\psi(g)|^2$ gives
\beq
 \int_{G} f(g) \log f(g) dg \leq \frac{n}{2} \log \left[\frac{C_G}{2\pi e n} \int_{G} \frac{1}{f} 
\|{\bf \tilde{X}} f\|^2 \right] dg \,\, \Longrightarrow \,\, -S \leq \frac{n}{2} \log \left[\frac{C_G}{2\pi e n} 
{\rm tr}(F) \right]
\label{logsobolieineq3}
\eeq
The rest is the same as for the case of $\IR^n$.

Starting with Gross's original form of log-Sobolev inequalities involving the heat kernel, the following
information-theoretic inequality results:
\vskip 0.1 true in
\noindent
{\bf Theorem 6.2}: The Kullback-Leibler divergence and Fisher-Information distance of any arbitrary pdf
and the heat kernel are related as
\beq
D_{KL}(f\, \| \, \rho_t) \leq \frac{c_G(t)}{2} D_{FI}(f\, \| \, \rho_t) 
\label{dfidkl}
\eeq
where in general given $f_1(g)$ and $f_2(g)$, 
\beq
D_{FI}(f_1 \, \| \, f_2) = \int_G \left\| \frac{1}{f_1} {\bf \tilde{X}} f_1 -  \frac{1}{f_2} {\bf \tilde{X}} f_2 \right\|^2 f_1 dg. 
\label{fidistdefgr}
\eeq
\vskip 0.1 true in
\noindent
{\bf Proof}: Starting with (\ref{logsobolieineq2}), let $\psi(g,t) = [\rho(g,t)]^{-\half} [f(g)]^{\half}$
where $f(g)$ is a pdf. Then
$$ \int_{G} |\psi(g,t)|^2 \rho(g,t) dg = \int_{G} f(g) dg = 1 $$
and so $\log \|\phi\|_{L^2(G,\rho_t)}^2 =0$, and we have
$$ \half \int_G f(g) \log \frac{f(g)}{\rho(g,t)} dg \leq \int_G \|{\bf \tilde{X}} ([\rho(g,t)]^{-\half} [f(g)]^{\half}) \|^2 \rho(g,t) dg. $$
By using the chain rule and product rule for differentiation,
$$ {\bf \tilde{X}} ([\rho(g,t)]^{-\half} [f(g)]^{\half}) = \half f^{-\half} {\bf \tilde{X}} f - \half f^{\half} \rho_t^{-1} {\bf \tilde{X}} \rho_t. $$
Substititution into the right hand side of (\ref{logsobolieineq2}) then gives (\ref{dfidkl}).

In the functional analysis community from which log-Sobolev inequalities emerged it is rarely, if ever,
stated in these terms. One exception is the work of Carlen \cite{carlen}, which addresses Theorem 6.1 for
the case of $G = \IR^n$. Moreover, the author has not found analogs of (\ref{logsobofish}) in the
context of Lie groups in the literature.

\section{The Entropy-Power Inequality (or Lack Thereof)}

One of the fundamental inequalities of information theory is the entropy power inequality
$$ N(f_1 * f_2) \geq N(f_1) + N(f_2)  $$
for any pdfs $f_1$ and $f_2$ on $\IR^n$ with $N(f_i)$ defined as in (\ref{logsobfingr}) for $C_{\IR^n} =1$.
This was first stated by Shannon \cite{shannon} together with a verification of the necessary conditions
for it to be true. This was followed up with proofs of sufficiency by Stam and Blachman \cite{blachman,stam}.
Without going into too many details, the key technical points of their proofs require two properties.
First,
$$ f_1 * \rho_{t_1} * f_2 * \rho_{t_2} = f_1 * f_2 * \rho_{t_1} * \rho_{t_2} $$
(which is not a problem in $\IR^n$ since convolution is commutative). Second, they also use a scaling
argument requiring that any pdf $f(\xx)$ that is scaled as $f_{s}(\xx) = s \cdot f(s \cdot \xx)$
will become the Dirac delta function as $s\rightarrow 0$.
That is not to say that these two properties are essential to proving the entropy power inequality,
but rather only that they are the properties that are used in the most familiar proofs.

However, there is somewhat of a conundrum because for compact Lie groups, the heat kernel $\rho_{t}(g)$
is a class function, and therefore satisfies the first condition. However, there is no natural way to rescale
on a compact Lie group (not even on the circle group, $SO(2)$). And in fact, it is easy to see that on
compact Lie groups the entropy power inequality does not hold. For example, the limiting distribution on
a compact Lie group is $\rho_{\infty} = 1$ with entropy $S(\rho_{\infty}) = 0$, and entropy power
$N(\rho_{\infty}) = 1$. Since $\rho_{\infty}* f = \rho_{\infty}$ for any pdf, $f$, we get $N(\rho_{\infty}* f) = 1
\ngeq 1 + N(f) $ since $N(f) > 0$ always.

On the other hand, it is possible for some groups to introduce a concept of scaling. For example, it is possible
to do this in the Heisenberg group, roughly speaking, because all coordinate directions extend to infinity.
Groups that admit a scaling property have been studied extensively \cite{follandstein}. However, whether the
heat equations on such groups yield solutions that are class functions then becomes an issue.
Regardless, for the groups of primary interest in engineering applications, i.e., the rotation and
rigid-body motion groups, the possibilities for an entropy power inequality appear to be pretty slim.

%Having said this, in most real-world problems, data measured on groups tends to be concentrated. And therefore
%inequalities that are globally valid may not be necessary to be useful in applied domains.
%
%**** holds locally for small diffusion ?? ****

\section{Conclusions}

By collecting and reinterpreting results relating to the study of diffusion processes, harmonic analysis,
and log-Sobolev inequalities on Lie groups, and merging these results with new definitions of covariance and
Fisher information, many inequalities of information theory were extended here to the context of probability
densities on unimodular Lie groups. In addition, the natural decomposition of groups into cosets, double cosets,
and the nesting of subgroups provides some inequalities that result from the Kullback-Leibler divergence of
probability densities on Lie groups. Some special inequalities related to finite groups were also provided.

While the emphasis of this paper was on the discovery of fundamental inequalities and the introduction of
Lie group concepts to the information theory audience, the motivation for this study originated with applications
in robotics and other areas. Though these applications were not explored here, references to the literature
pertaining to robot motion and image reconstruction were provided.

\vskip 0.1 true in
\noindent
{\bf Acknowledgments}

This work was performed under support from the NIH Grant
R01 GM075310 ``Group Theoretic Methods in Protein Structure
Determination.''

\end{document}